\documentclass{aa}  

\usepackage{graphicx}
\usepackage{xcolor}
\usepackage{txfonts}

\usepackage[]{hyperref}
\def\msun{\hbox{M$_\odot$}}

\begin{document}

   \title{A first GLIMPSE into star clusters populations across cosmic time}
   \titlerunning{Star clusters across cosmic time}

   \subtitle{}

   \author{A. Claeyssens
          \inst{1}\thanks{Corresponding author: Adélaïde Claeyssens:\\ \href{mailto:adelaide.claeyssens@univ-lyon1.fr}{adelaide.claeyssens@univ-lyon1.fr}},
          A. Adamo\inst{2},
          V. Kokorev\inst{3,4},   
          L. Furtak\inst{3,4},
          J. Richard\inst{1},
          B. Beauchesne\inst{5,6},
          M. Dessauges-Zavadsky\inst{7},
          \newline
          H. Atek\inst{8},
          J. Chisholm\inst{3,4},
          R. Endsley\inst{3,4},
          S. Fujimoto\inst{3},
          D. Korber\inst{7},
          R. Pan\inst{9},
          A. Saldana-Lopez\inst{2}, 
          D. Schaerer\inst{7}
          }
   \authorrunning{Claeyssens et al.}
   \institute{Univ Lyon, Univ Lyon1, Ens de Lyon, CNRS, CRAL UMR5574, F-69230, Saint-Genis-Laval, France
         \and
             Department of Astronomy, Oskar Klein Centre, Stockholm University, AlbaNova University Center, SE-106 91, Sweden\label{univstock}
         \and
             Department of Astronomy, The University of Texas at Austin, Austin, TX 78712, USA
         \and 
             Cosmic Frontier Center, The University of Texas at Austin, Austin, TX 78712, USA 
         \and
             Centre for Extragalactic Astronomy, Department of Physics, Durham University, South Road, Durham DH1 3LE, UK
         \and 
             Institute for Computational Cosmology, Department of Physics, Durham University, South Road, Durham DH1 3LE, UK
         \and 
             Department of Astronomy, University of Geneva, Chemin Pegasi 51, 1290 Versoix, Switzerland
         \and 
             Institut d'Astrophysique de Paris, CNRS, Sorbonne Universit\'e, 98bis Boulevard Arago, 75014, Paris, France
         \and 
             Department of Physics \& Astronomy, Tufts University, MA 02155, USA
             }

   \date{Received XXX accepted XXX}

 
  \abstract
  {We present the first sample of 222 high-redshift ($z>0.5$) star clusters, detected with JWST/NIRCam in 78 magnified galaxies from different galaxy cluster fields. The majority of the systems ($\sim 60 \%$) is observed in the very deep NIRCam observations of the cluster AbellS1063 (GLIMPSE program), showing the power that deep observations, combined with lensing, has to reveal these primordial stellar structures.
  We perform simultaneous size-flux estimates in all available NIRCam filters and spectral energy distribution (SED) fitting analysis to recover star cluster physical properties. All star cluster candidates have very high magnification ($\mu$>10). Star clusters and clumps show similar ages and redshift distributions, although noticeable differences are seen in their masses, sizes and stellar surface densities inherent to the lack of resolution in the latter group. We reconstruct the formation redshift of star clusters and find that the large majority of the observed star clusters show young ages ($<100$ Myr) and seems to form at cosmic noon (CN, $1<z<4$). A small sample of CN star clusters is about 1 Gyr old, these potential globular clusters have formed well within cosmic reionization. 
  Star clusters have stellar densities in the range $10^2$ to $10^6$ \msun/pc$^2$, with median values around $10^4$ \msun/pc$^2$. Their sizes and densities better overlap with those of nuclear star clusters in the local Universe. These intrinsic properties make high-z star clusters a viable channel to grow intermediate mass black holes. 
  We use Bayesian inference to make first direct measurement of the star cluster mass function at $z>1$, based on a subsample of 60 star clusters younger than 100 Myr and with masses above $2\times10^6$ \msun. The star cluster mass function is well described by a power-law with slope $\beta_{50\%} = -1.89^{+0.13}_{-0.12}$ suggesting that a power-law -2 function might already be in place in the distant Universe.
   
   }


   \maketitle
%

\section{Introduction}
\label{sec:intro}

Star clusters are among the fundamental building-blocks of galaxies, tracing their star formation history (SFH) across cosmic time (\citealt{PZ_2010review}). In the local Universe, the rich populations of young star clusters (YSCs) and globular clusters (GCs) offer a unique record of the conditions under which stars form across cosmic time (e.g. \citealt{adamo2020}). However, directly observing YSCs in higher redshift galaxies ($z>1$) remains challenging: even massive YSCs ($\sim10^5$ \msun) at these distances have luminosities below the detection limits (i.e., AB magnitude $>>29$ in JWST/NIRCam short wavelength filters) of the deepest pre-JWST imaging. Moreover their compact sizes ($<<20$ pc)  makes them unresolved or blend with other structures in most extragalactic observations beyond the Local Group because of the limited PSF size of current instrumentations. This is especially limiting at high redshift, where spatial resolution is limited to several 100s pc resolutions in the best cases. As a result, the physical processes that govern the formation, evolution, and survival of bound stellar systems in the early Universe remain largely unknown.

\noindent The advent of the James Webb Space Telescope (JWST) and its unprecedented combination of sensitivity and angular resolution in the near-infrared (NIR) has opened a new window onto star clusters at high redshift. The Near-Infrared Camera (NIRCam) instrument enables the detection of extremely faint, compact and distant galaxies (e.g., \citealt{Finkelstein2025,Carniani2024,Cox2025}), reaching rest-frame optical wavelengths beyond $z\sim1$. In strongly lensed fields, gravitational magnification boosts both the observed flux as well as the physical spatial resolution, allowing individual YSCs ($<20$ pc) and compact stellar clumps (several 10s of pc) to be resolved at cosmological distances (\citealt{claeyssens2023, vanzella2023, adamo2024, claeyssens2025, mowla2024, Ab2025}). These observations have revolutionized our view of stellar clump and star cluster formation in the distant Universe, revealing that massive (with typical stellar masses $\rm M_{*}$ ranging from $10^6$ to $10^8 \rm \ M_{\odot}$) and compact (with typical effective radii $\rm R_{eff}<100  \ pc$) stellar systems were already common within the first few hundred million years of the Universe \citep{bradac2025,nakane2025}. 

\noindent Recent single-case studies conducted in strongly magnified arcs provide the first key insights. In the Cosmic Gems arc at $z =9.6$, \citet{adamo2024} identified multiple $\sim1$ pc clusters with stellar masses of $\sim 10^6 \ \rm M_{\odot}$, extreme stellar surface densities (with $\Sigma_{*} \sim  \rm 10^5 \ M_{\odot}/pc^2$), and ages of only few 10s Myr, collectively accounting for the majority of the galaxy’s UV light. Further analyses of this system (\citealt{vanzella2025, messa2025}) revealed the clusters formed in a recent burst with high efficiency (30\% of the burst mass is in the 5 star clusters). 
Similarly, for the Firefly Sparkle, at $z=8.4$, \citet{mowla2024} measured that the galaxy mass is largely confined to $\sim$10 bound clusters contributing $>50$\% of the total stellar mass. At $z=6$, \citet{vanzella2023} measured in the highly magnified ($\mu>100$) Sunrise arc, high-density, young and massive star clusters dominating the light of the galaxy. Together, these studies establish that massive and bound star clusters could have formed efficiently and abundantly during the epoch of reionization, and might play a central role in the build-up of early galaxies and the production of ionizing photons. At lower redshift, \citet{Mowla2022} and  \citet{claeyssens2023} identified within the Sparkler galaxy at $z = 1.38$, 10 evolved (with ages $>1$ Gyr) and compact GCs consistent with having formed at $z \sim 7 - 11$, thus bridging present-day GCs to their high-redshift progenitors.

\noindent In parallel, recent high-resolution hydrodynamical simulation studies have significantly advanced our understanding of star cluster formation within the early Universe. The “SIEGE” (Simulation of Early-universe Galactic Environments) simulation, (\citealt{Pascale2023,Calura2025,Pascale2025}) demonstrates that compact star clusters with parsec-scale sizes naturally form in a cosmological context when sub-parsec physics and feedback from individual stars are included. 
With star-by-star hydrodynamical simulations of cluster formation within isolated dwarf-galaxy environments, \citet{lahen2025a} show that gravitationally bound star clusters can form with very compact initial sizes (half-mass radii of $\sim0.1 - 1$ pc) and very high stellar surface densities (up to $\sim 7\times 10^4 \rm \ M_{\odot}/pc^2$). \citet{mayer2025} performed a high-resolution ($\sim2$ pc) cosmological zoom-in simulations of gas-rich galaxies at $z\sim7 - 8$, finding that massive and bound stellar clumps form rapidly from gravitational instability into the disks of galaxies in high-density environments. These clumps quickly turn into compact star clusters with stellar masses ranging between $10^5$ and $10^8 \ \rm M_{\odot}$ and typical sizes of a few pc, reaching stellar surface densities above $10^5 \ \rm M_{\odot}/pc^2$. Similar results are recovered at higher redshift, by \citet{williams2025} using an AREPO-based cosmological simulation box designed to resolve star clusters at $z>12$ in a feedback-free regime. They recover star clusters stellar surface densities between $10^3$ and $10^6 \ \rm M_{\odot}/pc^2$. Strong Ly$\alpha$ feedback is emerging to possibly be a dominant early feedback process during the formation of star clusters, and impose a lower limit on the stellar surface densities of young, gravitationally bound star clusters \citep{Nebrin2025}. In spite of the challenges of combining several orders of magnitudes physical processes and millions particle, resulting in assumptions and compromises, it is encouraging that numerical approaches recover star cluster properties approaching recent JWST observations (\citealt{adamo2024, vanzella2023, mowla2024, claeyssens2023}). However, significant tensions also remain. 
By comparing six different simulations (\citealt{Renaud2017, Kruijssen2019a, Reina-Campos2022, Chen2024, DeLucia2024, Valenzuela2024}) of Milky Way-mass analogues which included star cluster formation and evolution, \citet{Valenzuela2025} show that in most of the simulations the GCs surviving till redshift 0 formed at cosmic noon (hereafter CN, defined as $1<z<4$). The same simulations only start to form massive star clusters below redshift 6-7, in tension with JWST observations that report detection of star cluster like structures up to redshift 11 \citep{bradac2025, nakane2025}. 

\noindent Key questions about the nature of high-redshift star clusters remain open: How early were the first star clusters able to form? How do they evolve with cosmic time? Are they GC progenitors? Strong observational constraints on the star cluster mass function, formation redshift distribution, and survivability (i.e., the age reached before to be disrupted or merge with another structure) of these compact systems as a function of redshift are much needed to inform and further progress on theory.
Addressing these questions requires a statistical collection of observed star clusters across a wide range of redshift and luminosities enabling comparison with model predictions and with the properties of local GCs.

\noindent In this paper, we present the first large sample of star clusters (222), detected at $z>0.5$, with intrinsic effective radii (R$_{\rm eff}<20$ pc). It combines homogenously re-analyses of stellar systems published in the literature and 145 newly identified star clusters in the JWST/NIRCam imaging of AbellS1063, a strongly lensed galaxy field observed by the GLIMPSE program \citep{Atek2025}. GLIMPSE combines the magnification power of a massive cluster with the deepest NIRCam observations acquired so far (reaching observed ABmag of 30.9 at 5$\sigma$) to probe galaxy evolution from the faintest ends and enabling the first demographics of compact stellar systems across cosmic time. 
We characterize their luminosities, sizes, and physical properties, and build the first direct measurement of a star cluster mass function at $z>1$. Using their observed redshift and derived ages we discuss their formation redshift. The paper is organised as follows: Section~\ref{sec:section2} presents the GLIMPSE data, Section~\ref{sec:section3}, the construction of the star cluster sample from different NIRCam datasets. Section~\ref{sec:section4} presents the physical properties of these objects. In Section~\ref{sec:section5} we discuss these results in the context of our current understanding of star clusters formation and evolution across cosmic time.

\section{The GLIMPSE data}
\label{sec:section2}

\subsection{JWST/NIRCam observations}
\label{sec:JWST_obs_GLIMPSE}

A detailed description of the JWST/NIRCam observations, data reduction, and source extraction is presented in details in the GLIMPSE survey paper (\citealt{Atek2025}). We briefly summarize here the main steps. The ultra-deep imaging from the public GLIMPSE survey (PID: 3293, PIs: H. Atek and J. Chisholm) targets the galaxy cluster AbellS1063 (hereafter AS1063, at $z=0.351$). AS1063 is one the highest-magnification region in the Hubble Frontier Fields (HFF, \citealt{Lotz2017}). The NIRCam data consist of 155h of science time observations, reaching unprecedented depths of $30.6 - 30.9$ mag across 7 broadband filters (F090W, F115W, F150W, F200W, F277W, F356W, F444W) and 2 medium-band filters (F410M, F480M). The lensing magnification reached within AS1063 allows us to probe intrinsically faint and compact sources (with intrinsic AB magnitudes $<31$, \citealt{Kokorev2025}) which would otherwise go undetected even in the deepest JWST blank field surveys. NIRCam data have been reduced following the procedure described in \citet{Endsley2024}. Crucial enhancements over the standard STScI pipeline have been implemented \citep[see][for details]{Kokorev2025} to improve the background subtraction and enable the mosaics to reach $0.3$ mag deeper across all filters. The final data-set achieved $5\sigma$ aperture-corrected nominal depths of $30.8 - 30.9$ mag across all broad-bands over $\rm D=0.2''$ apertures. The final images are drizzled onto $0.02$''/pixel grid for the short wavelength (SW) filters and $0.04$''/pixel for the long wavelength filters (LW).

\subsection{GLIMPSE galaxy catalogue}
\label{sec:source_extraction}

The detailed of the source extraction procedure to build the GLIMPSE galaxy catalogue is described in \citet{Kokorev2025}. The sources are detected using SExtractor (\citealt{Bertin1996}) and the flux densities are measured using PHOTULTILS (\citealt{Bradley2020}) within the established segments. Photometric redshift estimates ($\rm z_{phot}$) have been obtained with the Python version of EAZY (\citealt{Brammer2008}) using the BLUE SFHZ 13 model subset that contains redshift-dependent SFHs and dust attenuation values. 

\subsection{Lens models}
\label{sec:lens_models}

While AS1063 is a Frontier Field cluster (\citealt{Lotz2017}) and has already multiple robust lens models available (\citealt{Beauchesne2024,Richard2021}), we use here as a reference a new lens model that take in account JWST observations, presented in details in \citet{Atek2025}.This model is based on the parametric method by \citet{Zitrin2015}, that has already been successfully used to build strong lensing models for other galaxy clusters (e.g. \citealt{Pascale2022, Furtak2023}). The model is based on 75 multiple images from 28 individual sources, 24 of which have spectroscopic redshift (from MUSE data) and achieve a final image plane average RMS of 0.54'' (\citealt{Kokorev2025,Korber2025,Chemerynska2025,Atek2025}). This model is used as the reference lens model for the GLIMPSE data in our analysis and named M1.

\noindent Two additional lens models have been built based on different methods. 

\noindent Model M2 is based on the originally published mass model from \citet{Beauchesne2024} (also described in \citealt{Thai2023}), but uses only strong lensing positions as constraints. It however includes additional multiple images identified from JWST-GLIMPSE. Combined with the previous systems this parametric model is constrained with a total of 153 multiple images originating from 58 sources. The average image plane RMS is 0.82\arcsec.

\noindent Model M3 was presented in \citet{Beauchesne2025a,Beauchesne2025b}. In contrast to the other two models, it does not include new multiply-imaged systems from JWST images. However, it relies on more mass probes than strong lensing alone. It includes an intra-cluster gas component constrained by X-ray emission. Stellar kinematics of the BCG and cluster members serve as secondary total mass probes to constrain the inner core up to the last $5$~$\rm kpc$ and improve the mass estimation of individual cluster galaxies.

\section{Data Analysis}
\label{sec:section3}

\subsection{The GLIMPSE sample}
\label{sec:glimpse_sample}

\begin{figure*}
    \centering
	\includegraphics[width=15cm]{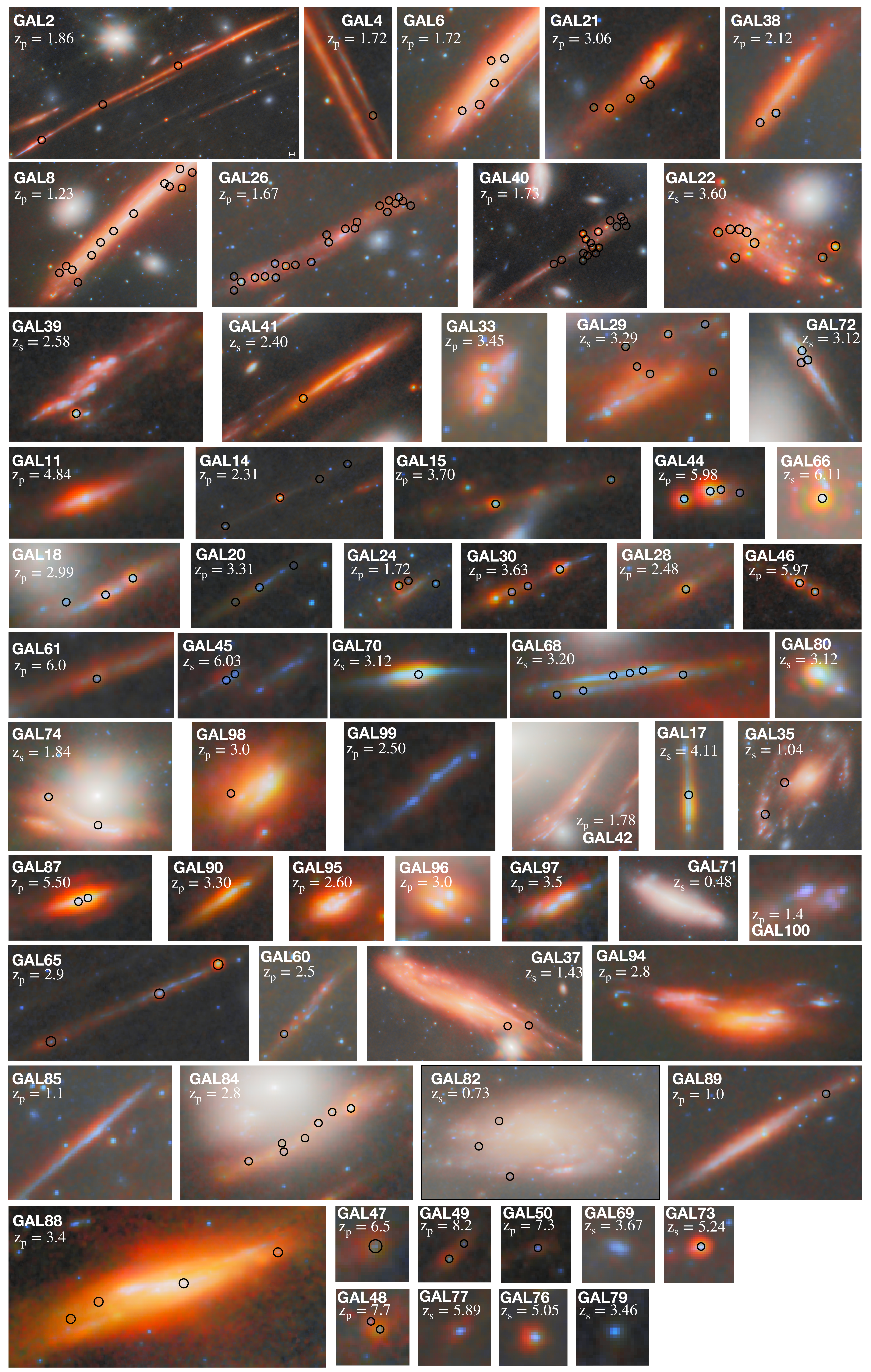}
    \caption{NIRCam composite color image of the 61 selected GLIMPSE galaxies. The image is a combination of all 7 broadband and 2 medium-band filters$^{1}$ . We show only one of the multiple images per system. The black circles indicate the position of the star cluster candidates ($\rm R_{eff}\leq 20$ pc). Redshift ($\rm z_p$ for photometric and $\rm z_s$ for spectroscopic redshift) is reported for each galaxy.}
    \label{fig:Mosaic_RGB}
\end{figure*}
\footnotetext[1]{\url{https://esawebb.org/images/potm2505a/}}

\subsubsection{Galaxy selection}
\label{sec:galaxy_selection}

We selected all galaxies with a robust photometric estimation (i.e., with a redshift uncertainty  $\Delta z < 0.5$) or, if available, a spectroscopic redshift (coming mostly from MUSE data, \citealt{Richard2021}) from the full GLIMPSE catalogue (\citealt{Kokorev2025}). In order to detect star cluster candidates (i.e., with intrinsic sizes $<20$ pc and intrinsic AB magnitude ranging from 25 to 33), we selected only galaxies with high magnifications ($\mu$ reaching at least 5 within each source, which corresponds to physical scale $<$ 100 pc in $z>1$ galaxies in NIRCam/F150W). We used magnification maps to estimate the maximum magnification within each object in the full source catalog. This selection resulted mainly in elongated arcs and multiple-image galaxies. Most of these galaxies are extended and splitted in multiple segments during the detection phase, resulting in multiple entries in the standard galaxy catalogue. In order to extract the full photometry for these galaxies, we defined for each of them an outer contour enclosing all the light of the galaxy. These contours were based on visual inspection of both short wavelength (SW) and long wavelength (LW) filters RGB images, the comparison between multiple images when possible, and the consistency of the photometric/spectroscopic redshift estimations of all the segments possibly belonging to the same system. 61 galaxies have been selected with these criteria (85 when including  multiple images). The final photometry was performed by applying an updated segmentation mask (contour) to the PSF matched images. The background for each target was estimated in close proximity of the system, to account for local variations. The uncertainty is a quadrature sum of the error array values contained within that segment. The selected galaxies are presented in the Figure~\ref{fig:Mosaic_RGB}. 

\subsubsection{Clump detection and photometry}
\label{sec:clumps_detection}

In this initial phase, the intrinsic R$_{\rm eff}$ of the compact stellar systems within galaxies were unknown so we refer to them as stellar clumps in this Section. Clump detection and photometry were performed using the methods and tools presented in \citet{claeyssens2025}. Clumps have been detected using Sextractor (through the python package sep, \citet{sep}) on the NIRCam filter F150W, selected because of its sharp and good image quality \citep[e.g.,][]{claeyssens2023, claeyssens2025}. 
Each galaxy has been visually inspected in all the available NIRCam observations, and clumps have been removed/added to the selection when needed. With this inspection, it has been ensured that all the visible compact and bright structures were detected. The final catalogue contains 489 clumps (693 detections when including multiple images of the same objects). The selected galaxies host between 1 and 42 clumps. 

\noindent Clump intrinsic sizes and fluxes have been estimated according to the methodology presented in \citealt{claeyssens2023} and \citealt{claeyssens2025}. To this effect, we derived the clump intrinsic size using as a reference the F150W images. The clumps have been modelled with an elliptical 2D Gaussian profile convolved with the instrumental PSF in the image plane (modelled from a stack of 5 bright, non-saturated and isolated stars). A 2D plane model of the immediate diffuse emission in proximity of the clump has been included to account for the spatially-varying local background emission from the host galaxy. The clump fit (based on the combination of a 2D gaussian model and a polynomial 2D surface as background) was performed on a cutout image ($9\times 9$ pixels) centred on each clump. If 2 clumps were separated by less than 4 pixels, they have been fitted together in a $13\times13$ pixels cutout.  The clumps were then fitted in the other bands by treating only the flux intensity and the local background as free parameter and fixing the shape of the light distribution to the intrinsic 2D gaussian best-model derived from the reference image convolved by the corresponding PSF of the given band. The position of the centre has been allowed to vary by 1 pixel maximum to account for drizzling and shifts due to different pixel scales among the data. 
According to the testing performed by \citet{Messa2019, Messa2022}, we assumed $0.4$ pixel as the minimum resolvable Gaussian axis standard deviation (std) in the F150W frame. If a clump was only resolved along the shear direction of magnification (i.e., the minor axis std is $<$ 0.4 pixel but the major axis std is $>$ 0.4 pixel) we estimated the size only based on the major axis value, assuming the clump is intrinsically circular in the source plane. If a clump was not resolved (i.e. the two axes std are $<$ 0.4 pixel), the size has been reported as upper limit at 0.4 pixel. To derive the intrinsic effective radius, $\rm R_{eff}$, we first estimated the radius of the circle having the same area of the ellipses describing the morphology of the clump, i.e., $\rm R_{cir} = \sqrt{x_{std} \times y_{std}}$. We assumed that $\rm R_{cir}$ is the standard deviation of a 2D circular Gaussian, and derived the observed PSF-deconvolved effective radius as $R_{eff,obs} = R_{cir} \times sqrt(2ln(2))$. When clumps were resolved in both direction, we derived their intrinsic effective radius $\rm R{eff,int}$ by dividing $R_{eff,obs}$ by $\sqrt{\mu}$, while when the clumps were unresolved in one or both direction, we directly divided $R_{eff,obs}$ by the tangential magnification $\rm \mu_T$. When clumps were unresolved in both directions, the $\rm R{eff,int}$ was estimated as an upper limit, derived as the half the full width half maximum  divided by the tangential magnification $\rm \mu_T$. In total, we have estimated sizes and fluxes measurements for 693 clumps (including multiple images) from the GLIMPSE galaxies.

\subsubsection{Contamination from intra-cluster globular clusters}
\label{sec:contamination_GC}

\begin{figure}
    \centering
	\includegraphics[width=8cm]{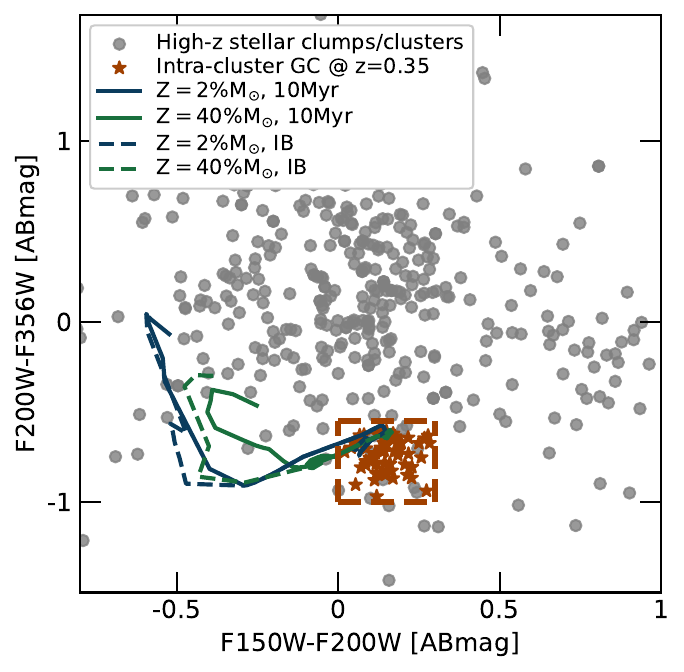}
    \caption{Color-color diagram (F150W-F200W vs. F200W-F356W) of the GLIMPSE high-redshift stellar clumps (grey points) and identified intra-cluster GCs (dark orange stars) at $z=0.35$. The blue and green lines show the Yggdrasil stellar evolutionary tracks based instantaneous burst (IB) and constant star formation (over 10Myr) models, with metallicity fixed at $\rm 2\% \ Z_{\odot}$ or $\rm 40\% \ Z_{\odot}$ and with a nebular covering fraction of 0.5 at $z=0.35$. The tracks start at 1 Myr and end at 7 Gyrs. The orange frame highlights the color-color location of the intra-cluster GCs. Stellar clumps located within this box, have been excluded from the final sample.  }
    \label{fig:intra_cluster_GCs}
\end{figure}

Giving the depth of the GLIMPSE NIRCam data, there is a considerable chance that intra-cluster GCs are detected across the strong lensing area. These systems present observed properties similar to those of high-redshift stellar clumps (i.e. compact sources) but can be identified through their distinct photometric colors. In order to identify potential interlopers within our sample, we selected 100 distinct intracluster GCs (identified visually from their PSF-shape and distinct blue color in a F277W/F356W/F444W RGB image), within the strong lensing area of the galaxy cluster, and we extracted their photometry in all the NIRCam bands (using the same method as for the high-redshift clumps). Intracluster GCs are all located in the same restricted region of a F150W-F200W vs. F200W-F356W color-color diagram (see Fig.~\ref{fig:intra_cluster_GCs}), within a 0.4 window in ABmag, corresponding to old and metal-poor stellar populations ($\rm -1 < F200W-F356W<-0.5$ and $\rm 0<F150W-F200W<0.3$), while the identified clumps populate the full color space. In order to minimise the contamination of our star cluster sample with intracluster GCs, we excluded all the clumps located within the area defined above (see Fig.~\ref{fig:intra_cluster_GCs}). In total we excluded 38 clumps from 25 different galaxies. Our final GLIMPSE sample is composed of 451 individual stellar clumps.

\subsection{The final GLIMPSE clump/star cluster sample}
\label{sec:glimpse_sample}

\begin{figure}
	\includegraphics[width=9cm]{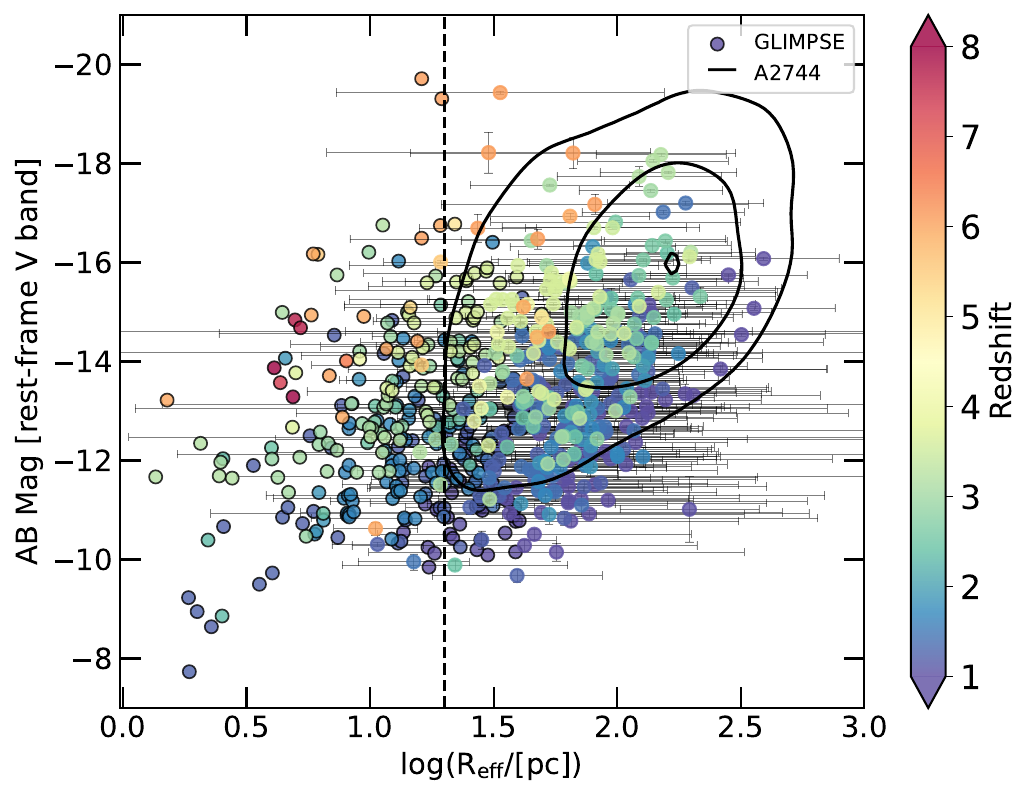}
    \caption{GLIMPSE stellar clumps absolute AB magnitude (corresponding to the V band, rest-frame) as a function of their effective radius. The points are color-coded in redshift. The black contours highlight the distribution of Abell2744 stellar clumps in the same plane (\citealt{claeyssens2025}, the three contours represent 10, 50 and 99\% of the sample). The vertical black line shows the star cluster size threshold (fixed at 20 pc for this study). The symbols circled in black indicate systems that have only a size upperlimit. Thanks to very deed observations and strong lensing magnification, the GLIMPSE sample enables us to detect fainter and smaller clumps than in Abell2744 NIRCam observations.}
    \label{fig:Mag_Reff}
\end{figure}

\noindent The Figure~\ref{fig:Mag_Reff} showcases the GLIMPSE clumps absolute AB magnitude (measured in the V band, rest-frame) as a function of their effective radius and compared to the properties of the stellar clumps extracted from A2744 NIRCam data (\citealt{claeyssens2025}). Thanks to the depth of the GLIMPSE data, combined to very high magnifications, we reach very faint magnitudes ($>-12$) for individual clumps. This has led to the detection of 5 times more star cluster candidates than in A2744 data (cf Table 1). 

For the rest of the study, we consider star cluster candidates all clumps with an effective radius value (or upper limit on the effective radius) smaller than 20 pc (dashed vertical line in Figure~\ref{fig:Mag_Reff}). The choice of this size limit is arbitrary. In the local universe, star clusters have radius distributions peaked around 2--3 pc \citep[e.g.][]{brown2021}. However, we stress that due to the upper limits and errors, a stricter limit would exclude genuine star cluster candidates. We discuss the impact of this criterion on the recovered physical properties in the Section~\ref{sec:section5}. This selection resulted in 145 star cluster candidates only from the GLIMPSE region, coming from 44 galaxies between $z=0.5$ and $z=8.2$, making it the largest sample of cluster candidates collected so far. Galaxies host between 1 and 38 star cluster candidates each. All the star clusters candidates from the GLIMPSE sample are indicated as black circles in Figure~\ref{fig:Mosaic_RGB}.

\begin{figure*}
    \centering
	\includegraphics[width=18cm]{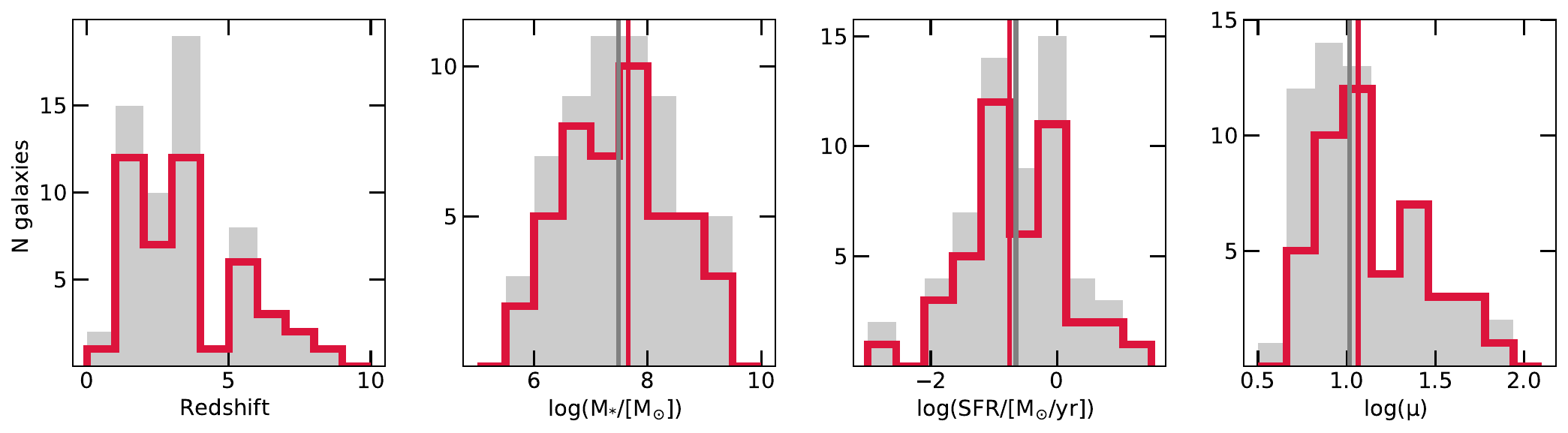}
    \caption{GLIMPSE galaxy properties, from left to right: stellar mass ($\rm M_{*}$), SFR (estimated over the last 10 Myr) and median magnifications ($\mu$) extracted from the reference lens model. The stellar mass and SFR are obtained from \texttt{Bagpipes} SED fitting using a non-parametric SFH (\citealt{Leja2019}) as described in the Section~\ref{sec:glimpse_sample}. The grey distributions represent the total GLIMPSE galaxy sample hosting clumps, while the red ones highlight the galaxies which host star cluster candidates. Vertical lines shows the median values of each distribution.}
    \label{fig:prop_galaxies}
\end{figure*}

\subsection{Literature samples}
\label{sec:literature_samples}

In order to increase our sample, we leverage the literature and include stellar clumps and clusters from other lensing fields (cf Table~\ref{tab:table_samples}): Abell2744 \citep{claeyssens2025}, SMACS0723 (\citealt{Mowla2022,claeyssens2023}, including the Sparkler galaxy at $z=1.37$), SPT0615 (The cosmic Gems arc at $z=9.6$, \citealt{adamo2024, vanzella2025, messa2025}, WHL0137 (The Sunrise arc at $z=6$, \citealt{vanzella2023}), MACSJ1423 (the Firefly Sparkle galaxy at $z=8.3$, \citealt{mowla2024}) based on the availability of the clump photometry and magnification information extracted in a similar way as in this work. Four of these datasets (Abell2744, SMACS0723, the Sunrise arc, and the Cosmic Gems arc) have been processed (for clump detection and photometry extraction) with the same method as used for the GLIMPSE sample, presented in Section~\ref{sec:section3}. The Firefly Sparkle clusters photometry and size have been extracted by \citet{mowla2024}.

\subsubsection{SMACS0723}
\label{sec:smacs0723}

\citet{claeyssens2023} built the very first sample of JWST lensed clumps. Among the 223 clumps studied in this region, 32 (coming from 8 different galaxies) have an effective radius smaller than 20 pc. These sizes are based on the same catalogue used in \citet{claeyssens2023}, updated with the latest version of the lens model published in \citet{Mahler2023}. A comparison with the values published in \citealt{claeyssens2023} is presented in the Appendix B. Among these star clusters, 5 come from the Sparkler galaxy ($\rm z_{spec}=1.34$), where GC candidates with ages raging from few 100s Myr to few Gyr, have been identified in the outskirts of the galaxy (\citealt{Mowla2022, adamo2023}). 

\subsubsection{Abell2744}
\label{sec:abell2744}

\citet{claeyssens2025} have gathered a sample of 1956 lensed clumps from 476 galaxies, at $0.7<z<10$, from the NIRCam observations of the lensing cluster A2744. We selected as star cluster candidates only clumps with effective radii value or upper limit smaller than 20 pc. Among the full sample, 34 clumps (i.e., 2\%), coming from 23 galaxies, have been selected.

\subsubsection{The Cosmic Gems}
\label{sec:cosmic_gems}

\citet{adamo2024} characterised five YSCs within a $\rm z_{spec}=9.6$ galaxy strongly lensed by the cluster SPT0615. These star clusters have effective radii smaller than 2 pc and magnifications ranging from 50 to 400. 

\subsubsection{The Sunrise arc}
\label{sec:sunrise}

\citet{vanzella2023} identified six YSCs within the Sunrise arc at $\rm z_{spec}=6$.  They have sizes between 1.4 and 25 pc and stellar mass ranging from 1 to $\rm 10 \times 10^6 \ M_{\odot}$. The four youngest star clusters (with ages $<$ 6 Myr) show evidence of prominent $\rm H_{\beta}+$[O {\sc iii}] emission and are hosted in a 200 pc sized star-forming complex. The survival of some of the clusters would make them the progenitors of massive and relatively metal-poor GCs in the local Universe. We included in our final star cluster sample the five cluster with size $<20$ pc.

\subsubsection{The Firefly Sparkle}
\label{sec:firefly_sparkle}

\citet{mowla2024} identified 10 star cluster candidates in the Firefly Sparkle at $\rm z_{spec}=8.3$ with upper limits on sizes (defined as half-light radius) $< 7$ pc and magnifications ranging from 15 to 26. The photometry of these systems has been performed using GALFIT in a similar way as done for all other samples included in in this study. 
\subsection{Retrieving physical properties}

\subsubsection{SED fitting of GLIMPSE galaxies}

\noindent  The galaxy properties have been obtained using the non-parametrized continuity SFH model by \citet{Leja2019}, as implemented in \texttt{Bagpipes} (\citealt{Carnall2019}).  The age bins, counted in Myr backwards from the time of observation, are: 10, 30, 50, 100, 200, 400, 600, 800, 1000, 1500, 2000, 3000, 5000, 7000, 8000, 10500. Depending on the redshift of the galaxies, we adjust between 10 and 14 age bins for 16 constraints (HST and JWST/NIRCam filters). This age bin distribution allows to constrain both a very recent burst of star formation ($<100$ Myr) and an older star formation episode ($>300$ Myr). The Figure~\ref{fig:prop_galaxies} shows the GLIMPSE galaxy properties (stellar mass $\rm M_{*}$, SFR and median magnification $\mu$ over the galaxies). The galaxies hosting star cluster candidates show similar stellar mass and SFR distributions than the full sample (cf Figure~\ref{fig:prop_galaxies}). 90\% of the galaxies with $\mu>15$ exhibit star cluster candidates.

\subsubsection{Star cluster SED fitting}
\label{sec:star_cluster_SED}

The total sample includes 222 individual star clusters detected in 78 different galaxies at $z=0.5 - 9.6$, with 65\% of them coming from the GLIMPSE sample. 
To further homogenise the star cluster sample we perform the same SED fitting analysis to all of them, including the systems published in the literature, with the fitting code\texttt{Bagpipes}(Bayesian Analysis of Galaxies for Physical Inference and Parameter EStimation, \citet{Carnall2019}) and BPASS v2.2.1 SED templates (\citealt{Stanway2018}), using the fiducial BPASS IMF with maximum stellar mass of 300 $\rm M_{\odot}$ and a high-mass slope similar to \citep{Kroupa2001}. Exponential declining SFHs have been used with different decline timescales, $\tau$ (1 and 10 Myr). For the two SFHs, the free parameters are the age ($\rm T_0$), with a uniform prior between the redshift of the source and the age of the Universe, the stellar mass formed ($\rm M_{formed}$), with a uniform prior between $\rm 10^{3}$ and $\rm 10^{15} \ M_{\odot}$, the metallicity ($\rm Z$), with a logarithmic prior between $\rm 10^{-3}$ and $1.5 \rm \ Z_{\odot}$, the dust content ($\rm A_V$) with a uniform linear prior between 0 and 1 (assuming a Calzetti dust law), and the ionisation parameter (log(U)) with a uniform linear prior between $-3$ and $-1$. As motivated in \citealt{claeyssens2023, claeyssens2025,adamo2024, vanzella2023} we assumed such short burst to approximate the condition that star clusters are typically assumed to be close to a single stellar population. We stress that the two SFHs produce similar physical parameter distributions. Only the ages are affected by the choice of duration of the burst. The SFH model with $\tau=1$ Myr, reproducing a single burst of star formation, is used as reference for the analysis, similarly done for YSCs in local galaxies (see review by \citealt{adamo2020}).  For the three literature samples analysed with different SED fitting codes (SMACS0723, the Sunrise arc and the Firefly Sparkler), we show a comparison of the extracted physical parameters (stellar mass, age and stellar mass surface density) in the Appendix B. 

\noindent The \texttt{Bagpipes}SED fitting outputs corrected with the $\mu$ factor of each region, produced the stellar mass ($\rm M_{*}$) and stellar mass surface density ($\rm \Sigma_{M_*}$), SFR and SFR surface density ($\rm \Sigma_{SFR}$), effective radius ($\rm R_{eff}$), mass-weighted ages ($\rm age$) of all the stellar clumps and star clusters, presented in the Figure~\ref{fig:histo_sample}. 

\begin{table*} 
\begin{tabular}{llllll} 
Field & Reference & Nb Clumps & Nb clusters & $\rm z_{clusters}$ & NIRCam filters \\  

\hline 
\hline 

AbellS1063 & This work & 451 & 145 & 0.5-8.2 & F090W, F115W, F150W, F200W, F277W\\ 
 (GLIMPSE)& & & & &  F356W, F444W, F410M, F480M\\ 
 \hline 

Abell2744 & \citealt{claeyssens2025} & 1920 & 25 & 1-8 & F070W, F090W, F115W, F150W, F200W\\ 
 & & & & &  F277W, F356W, F444W, F140M, F162M\\ 
& & & & & F182M, F210M, F250M, F300M, F335M\\
& & & & &  F360M, F410M, F430M, F460M, F480M\\ 
\hline 

SMACS0723 & \citealt{claeyssens2023} & 148 & 32 & 1-8.5 & F090W,  F150W, F200W, F277W, F356W\\ 
 & & & & &  F444W\\ 
 \hline 

WHL0137 & \citealt{vanzella2023} & 6 & 5 & 6 & F090W, F115W,  F150W, F200W, F277W\\ 
 (The Sunrise arc) & & & & &  F356W, F444W, F410M\\ 
 \hline 

SPT0615 & \citealt{adamo2024} & 5 & 5 & 9.6 &   F150W, F200W, F277W, F356W, F444W\\\ 
(The Cosmic Gems) & & & & &  F410M\\ 
\hline 

MACSJ1423 & \citealt{mowla2024} & 10 & 10 & 8.3 & F090W, F115W,  F150W, F150WN, F200W\\ 
(The Firefly Sparkle) & & & & & F200WN, F277W, F356W, F410M, F444W\\ 

\hline
\end{tabular} 
\caption{Summary of the different JWST/NIRCam samples of high-redshift star clusters used in this work. We list for each sample the cluster field, the reference paper, the total number of identified clumps, the total number of star cluster candidates selected (i.e., with effective radius $<20$ pc), the redshift range of the star cluster candidates and the NIRCam filter observations available.}
\label{tab:table_samples} 
\end{table*}


\section{Results}
\label{sec:section4}

\subsection{The first sample of high-redshift star clusters}
\label{sec:sample_star_clusters}

For the remaining of the article, we will focus on the combined GLIMPSE and literature sample. In total, the sample consists of 222 star cluster candidates (defined as stellar clumps with an effective radius smaller than 20 pc), coming from six different lensing fields and 78 different galaxies. Among the total sample, 98 (44\%) clusters are PSF-like point source detections and thus have an upper limit size measurements.

\begin{figure*}
	\includegraphics[width=18cm]{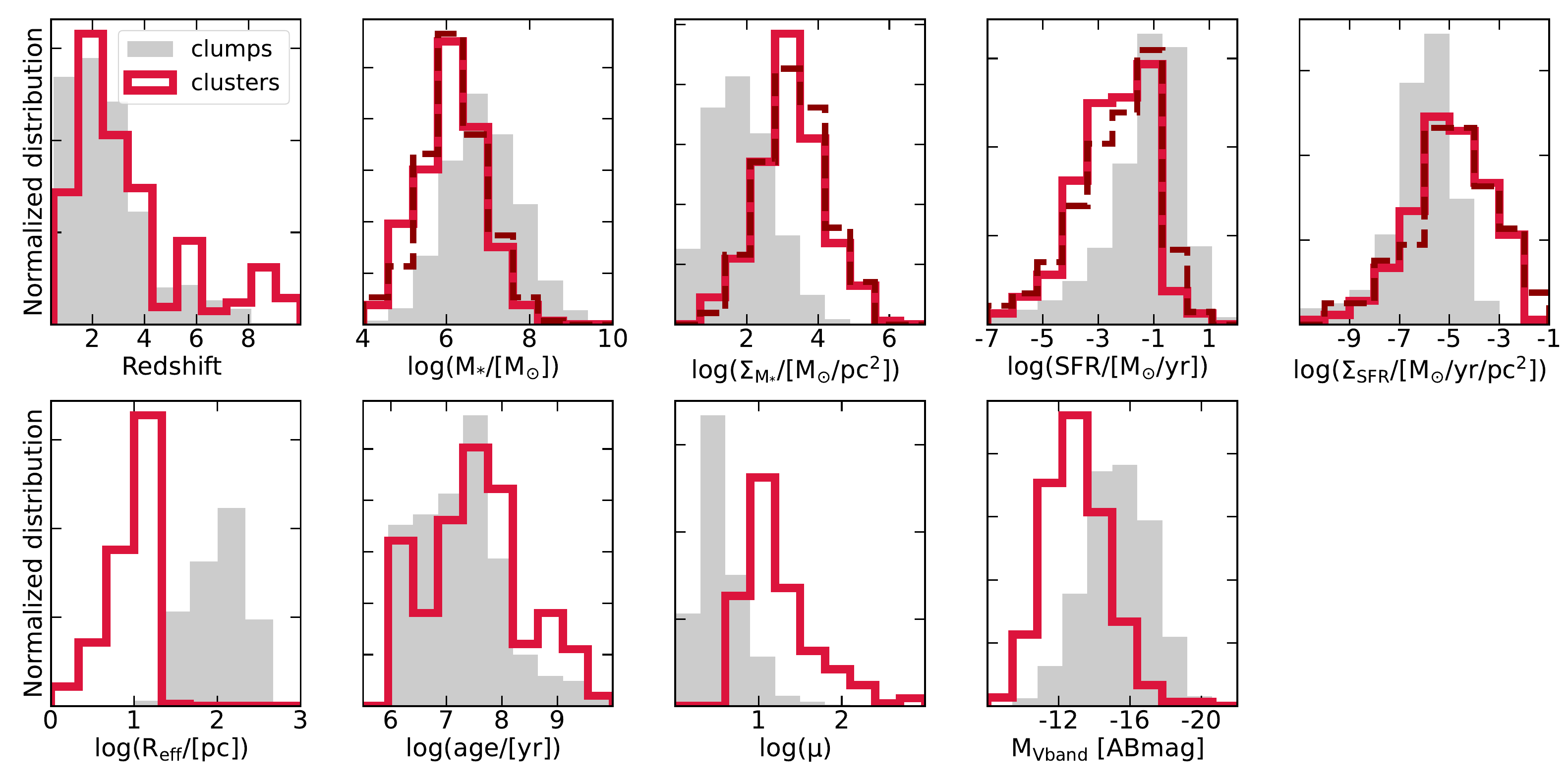}
    \caption{Properties of the JWST high-redshift stellar clumps (in grey, $\rm R_{eff}>20$ pc) and star clusters (in red, $\rm R_{eff}<20$ pc)). The distributions include the GLIMPSE sample as well as the 5 literature samples (see Section~\ref{sec:section3} and Table 1). The top row shows, from left to right, the redshift, the stellar mass ($\rm M_{*}$), the stellar mass surface density ($\rm \Sigma_{M_{*}}$), SFR and the SFR surface density ($\rm \Sigma_{SFR}$). The bottom row shows, from left to right, the effective radius ($\rm R_{eff}$), mass-weighted age, lensing magnification ($\mu$) and V band absolute magnitude ($\rm M_{Vband}$). The parameters obtained from SED fitting for the star clusters are measured with two different SFHs with exponential decline with $\tau=1 \ \rm Myr$ (in dark red, dashed) and $\tau=10 \ \rm Myr$ (in red). The parameters obtained from SED fitting for the stellar clumps are only measured using a SFH with exponential decline profile and $\tau=10 \ \rm Myr$ (grey). The histograms are not corrected for completeness.}
    \label{fig:histo_sample}
\end{figure*}

In Figure~\ref{fig:histo_sample}, we compare clumps with effective radii $>$ 20 pc vs. star cluster candidates.  Their redshift ranges from $z=0.5$ to $z=9.6$, with 75\% of them being detected at $z<4$. We notice a peak of star clusters at redshift $>8$ which is driven by including preferentially star cluster detection from the literature. Star cluster mass distribution covers from $10^4 \ \rm M_{\odot}$ to $10^8 \ \rm M_{\odot}$ with stellar mass surface density between $10^{1.5} \ \rm M_{\odot}/pc^2$ and $10^{6} \ \rm M_{\odot}/pc^2$. Stellar clumps easily reach larger stellar masses but their surface stellar densities are significantly lower because of their larger sizes (bottom left panel).
Star cluster SFRs are, on average, lower than stellar clumps, with 95\% of them having SFR below 1 $\rm M_{\odot}/yr$ while stellar clumps can reach 10s $\rm M_{\odot}/yr$ (cf Figure~\ref{fig:histo_sample}). However, star clusters reach very high SFR surface densities, up to $10^{-1} \ \rm M_{\odot}/pc^2/yr$. Star clusters exhibit a similar mass-weighted age distribution as stellar clumps, ranging between 1Myr to few Gyrs. The secondary peak in the star cluster age distribution is dominated by bona-fide GCs from the Sparkler and other detected in the GLIMPSE field (see next section). As expected, star clusters have the highest magnifications ($\mu>5$) and faintest (peaking at $\rm M_{VBand}>-12$) objects. By comparing the solid red versus the darkred dashed distributions we see that the star cluster physical properties produced by the two SED runs with different SFH assumptions (cf Section~\ref{sec:section3}) do not show significant deviations. 

\noindent Because the star cluster selection is only based on the clumps size it creates a  strong dependency on the reliability of the reference lens model. We tested this dependency  using the GLIMPSE sample as a test-bench. In the Appendix B (cf. Figure~\ref{fig:comp_models}) we compare the recovered star cluster magnifications ($\mu$) among the different models (presented in Section~\ref{sec:lens_models}). While recovered star cluster magnifications are consistent between the three models (cf. Figure~\ref{fig:comp_models}), a dozen of systems exhibit significant variations. They are all located very close to the critical line, subjected to strong magnification gradient, which is difficult to constrain explaining the strong variations for these systems ($\mu>15$ in all). When different lens models are used the resulting number of star clusters changes only slightly: 145 star clusters with the model M1, 140 with the model M2 and 152 with the model M3. The star cluster stellar masses and stellar mass surface densities distributions obtained from the three models (cf Figure~\ref{fig:comp_models}, middle and right panels) are very similar.

\noindent These results are similar to the trends reported in the literature sample included in this study by \citet{vanzella2023},  \citet{adamo2024}, and \citealt{claeyssens2025} which recover consistent values between the different lens models, in particular for the clump/star cluster sizes and stellar masses.

\subsection{On the formation of star clusters}
\label{sec:formation_star_clusters}

\begin{figure}
	\includegraphics[width=9cm]{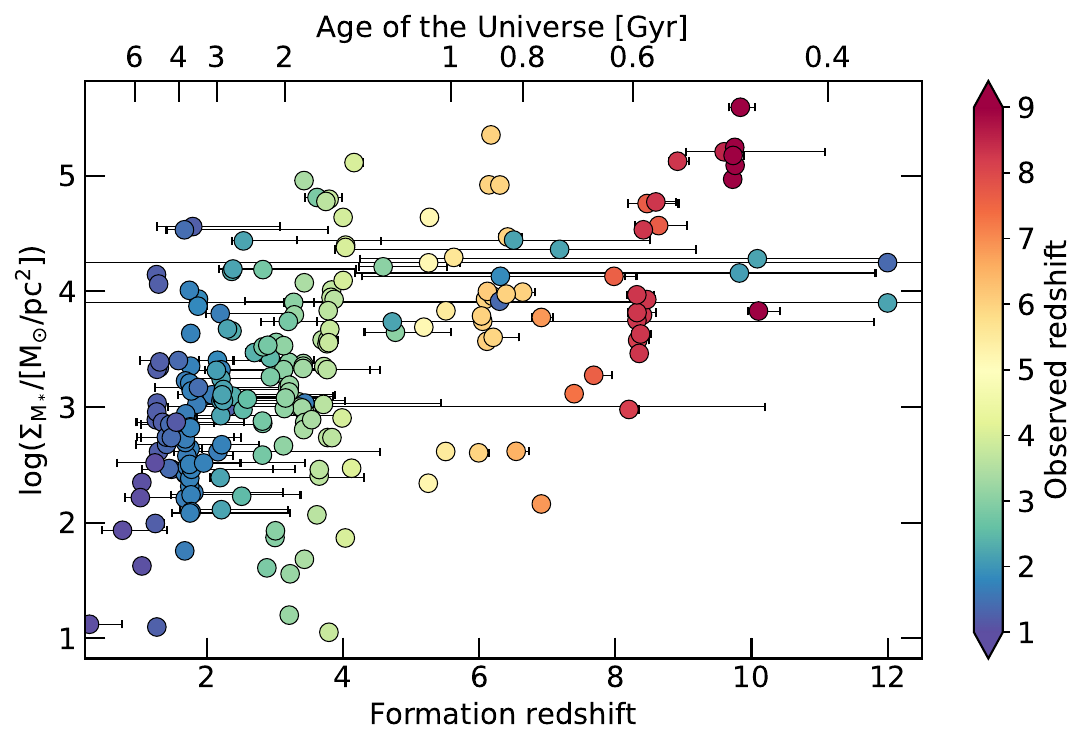}
    \caption{Star cluster stellar mass surface density ($\rm \Sigma_{M_{*}}$) vs. star cluster formation redshift. The points are colour-coded with the observed redshift. While a couple of CN observed star clusters seems to have formed before $z=5$, the large majority of the observed star clusters have formed during CN.}
    \label{fig:formation_redshift}
\end{figure}

Using the recovered star clusters ages and the redshift at which they are observed, we can recover their formation redshift: i.e., the time at which they formed. The formation redshift is defined the redshift when star formation starts according to the best-fit SFH obtained from BAGPIPES. As we use an exponential decline SFH for the clusters, this corresponds to the redshift of the unique peak of star formation. Figure~\ref{fig:formation_redshift} shows the formation redshift with respect to the star cluster stellar mass surface density and the observed redshift. We find that most of the star clusters in our sample are both detected and formed below $z<4$ as expected since that encompasses also the longest age span. A small fraction of star clusters (11 in total) detected at CN have formation redshift $>4$, with 5 being GC candidates from the Sparkler galaxy and 3 newly detected in the GLIMPSE field. The GC candidates have stellar mass and stellar mass surface densities ($>10^{3.5} \rm \ M_{\odot}/pc^2$) comparable to their young counterparts detected and formed at similar times ($z>6$), indicating that some of these clusters could survive at lower redshift. 
Figure~\ref{fig:formation_redshift} shows the star clusters stellar mass surface density versus their formation redshift. The points are colour-coded with the observed redshift of the star clusters. We see that we detect very few low surface density star clusters (with $\rm \Sigma_{*} < 10^{3} \ M_{\odot}/pc^2$) at $z>4$ (only 6 objects) while numerous low-density larger clumps are detected (cf. Figure~\ref{fig:histo_sample}). This lack of low density clusters at higher redshift is probably produced by completeness effect, as it is more difficult to detect compact and faint objects without very high magnification at larger distances. The bulk of young and massive star clusters at $1<z<4$ (cf Figure~\ref{fig:formation_redshift}) suggests an active cluster formation at CN which is consistent with recent simulations results (\citealt{Valenzuela2025}). We discuss this further in Section~\ref{sec:section5}. 

\subsection{Density--size distributions}
\label{sec:density_size}

\begin{figure*}

	\includegraphics[width=9cm]{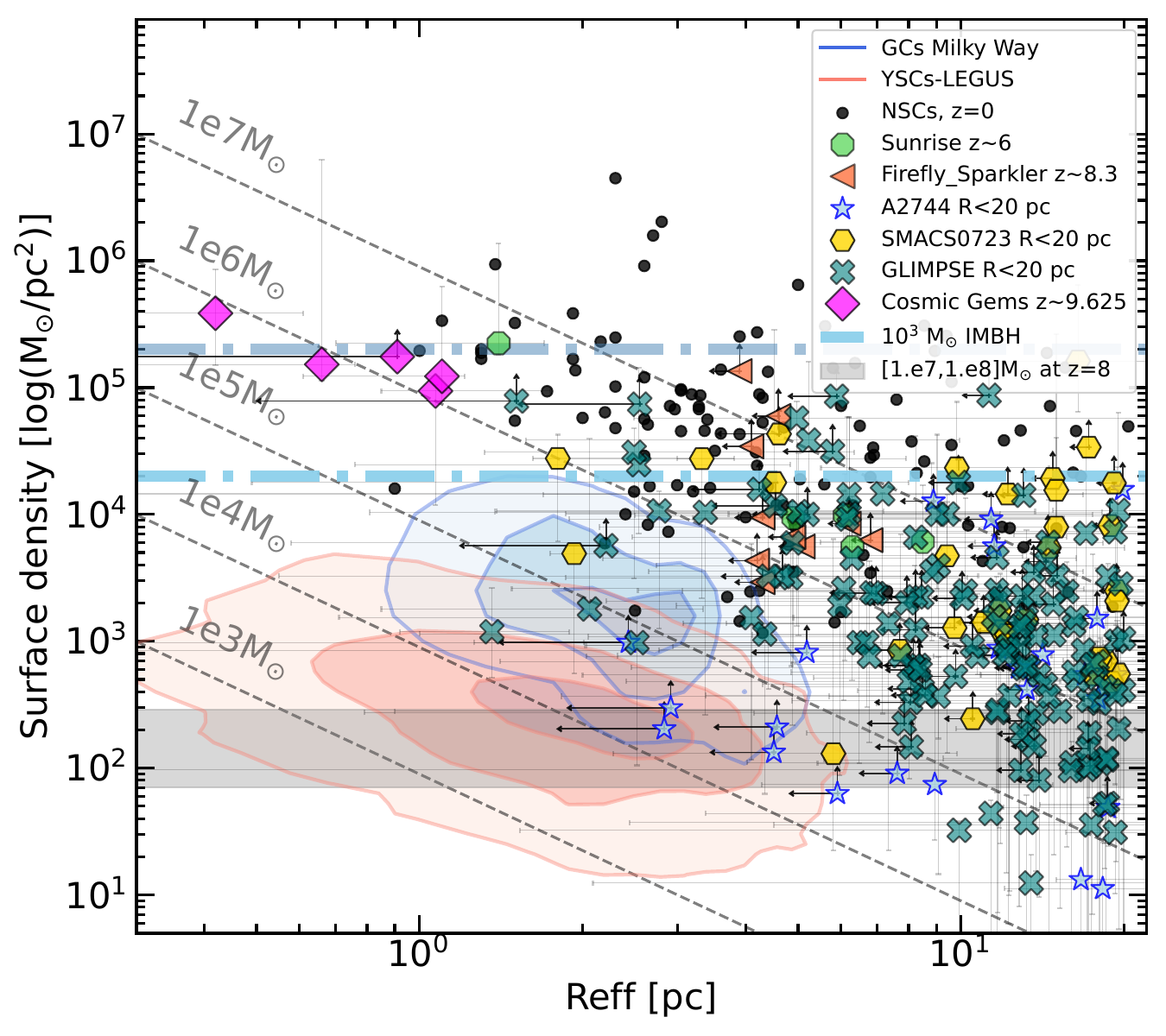}
  	\includegraphics[width=9cm]{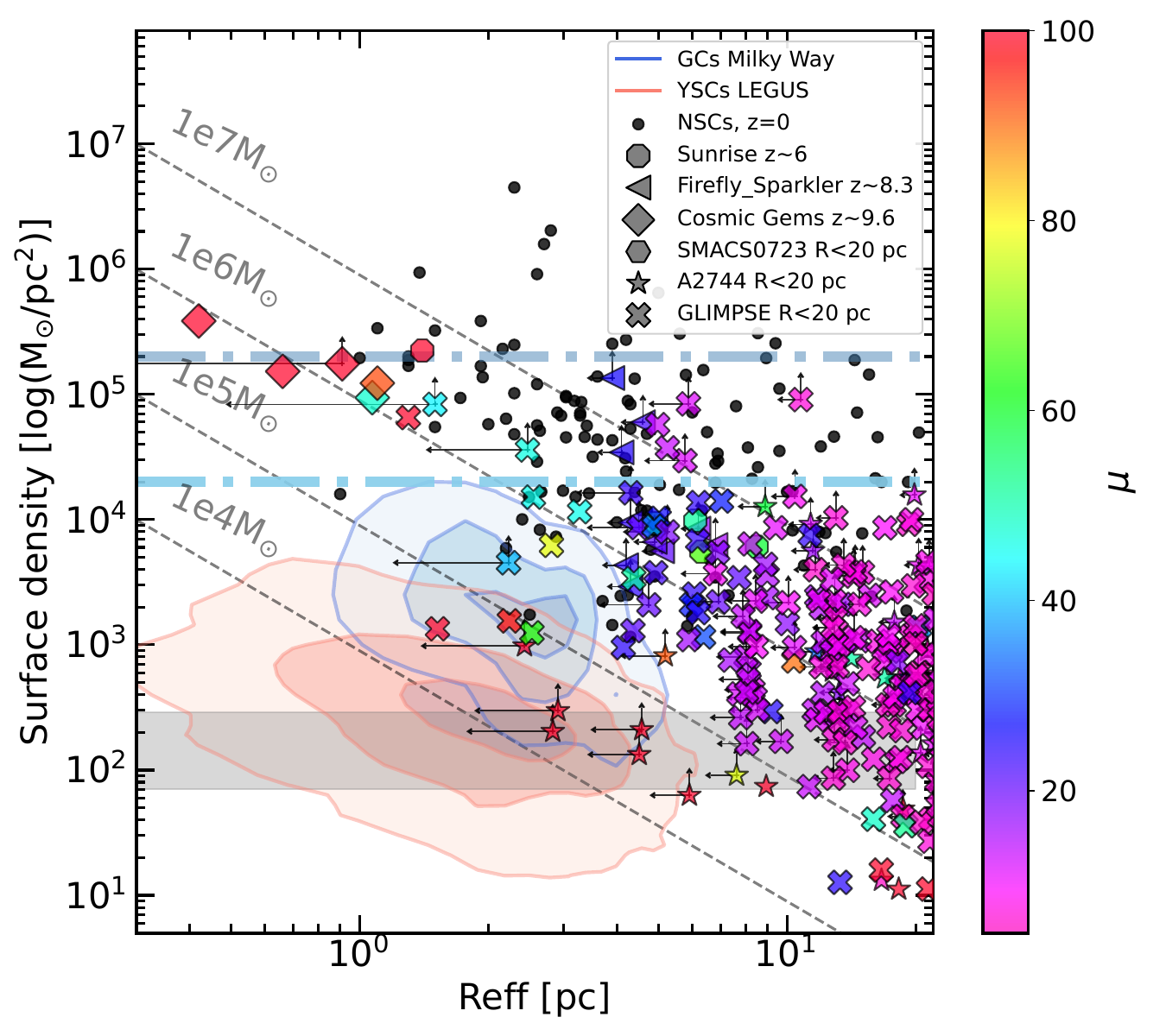}\\
    \includegraphics[width=9cm]{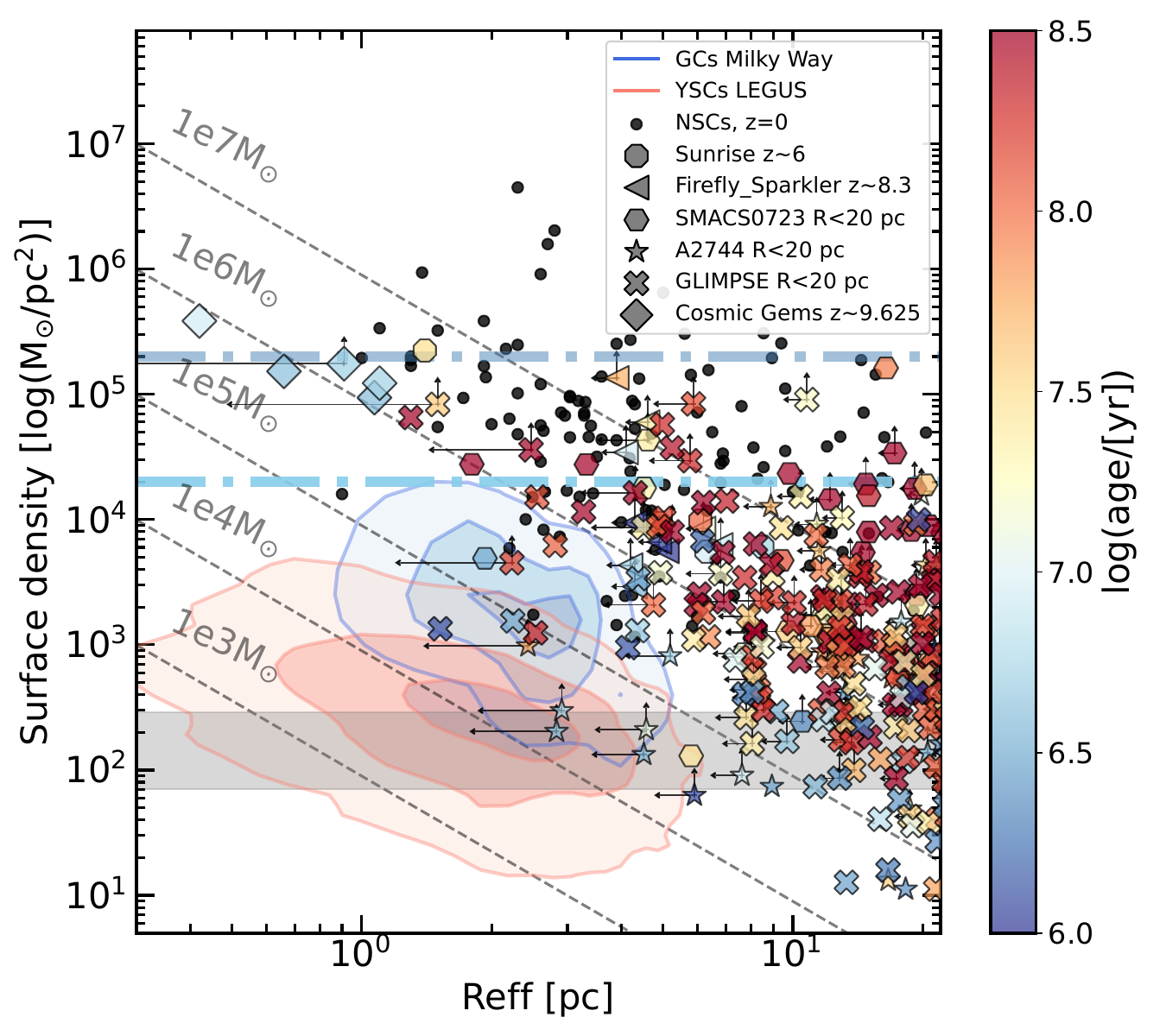}
     \includegraphics[width=9cm]{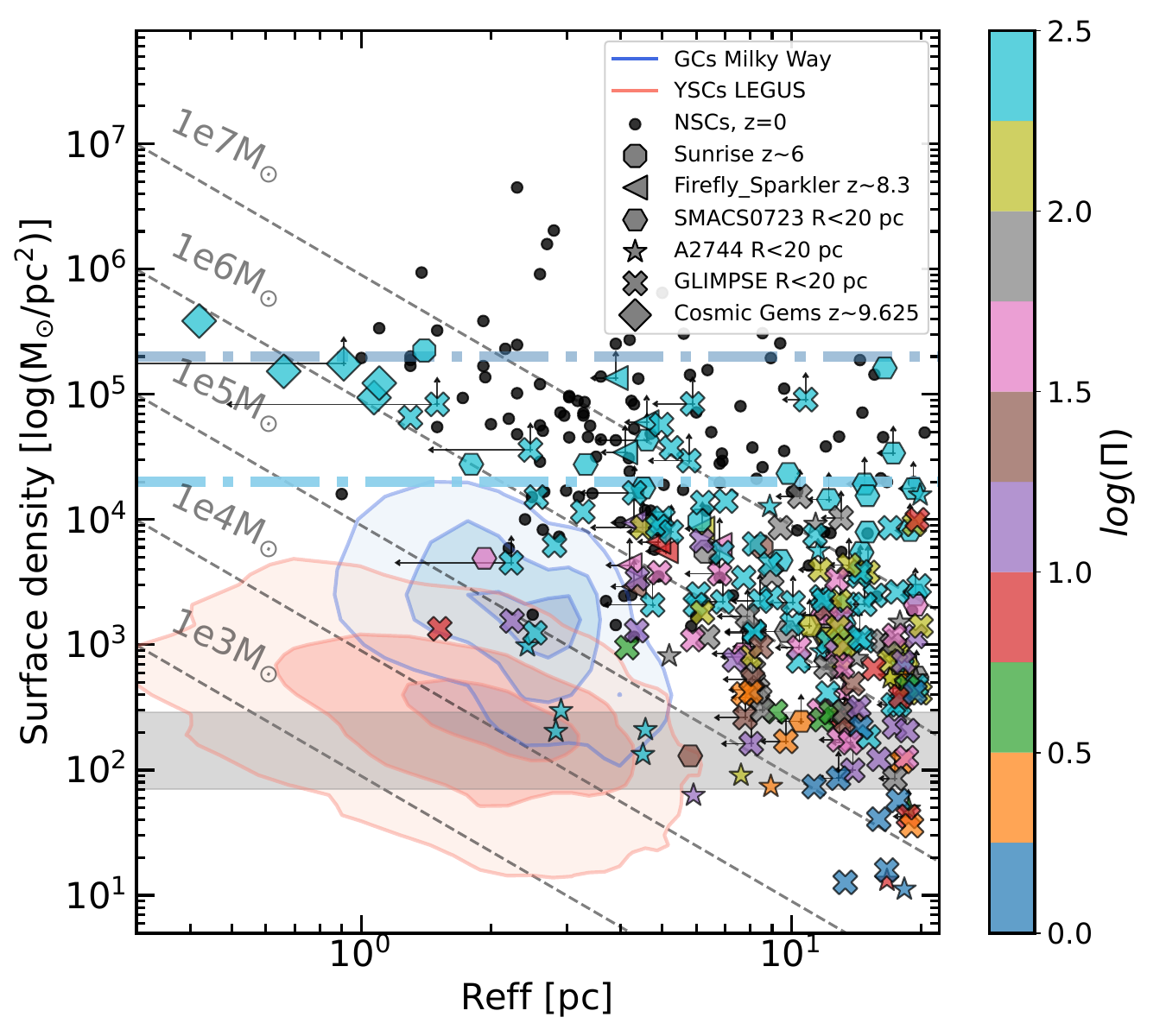} 
    \caption{Star cluster stellar surface density vs. R$_{eff}$ of the star cluster candidates in GLIMPSE and literature (see Section~\ref{sec:literature_samples}, colored symbols). Coral contours show the space occupied by YSCs \citep{brown2021} in the local Universe, while light-blue contours delimitate the space occupied by Milky Way GCs \citep{baumgardt2018}. The horizontal grey band show the average stellar density of an average  galaxy in the mass interval $10^7 - 10^8$\msun\ at $z=8$ extrapolated from the \cite{Morishita2024} relation. Black dots show the typical stellar densities of nuclear star clusters measured in local galaxies \citep{Neumayer2020}. The two thick dot-dashed lines shows the critical stellar densities which leads to IMBH formation above $10^3$ \msun\, at metallicities below 10\% solar abundances \citep{Rantala2025, Rantala2026}. The top left plot includes bootstrapped errors which are omitted for clarity in the other panels. We highlight in the other panels the magnifications (top right), ages (bottom left), and gravitational boundedness state of the star cluster candidates.}
    \label{fig:density}
\end{figure*}

We use the measured R$_{\rm eff}$ and derived stellar surface densities to investigate how high-z star clusters compare to YSCs, GCs and nuclear star clusters (NSCs) in the local universe. These comparisons are quite interesting since they could indicate the physical properties of the potential progenitors of GCs and NSCs. Moreover, they reveal how the different physical conditions under which star formation operates in high-z vs. local galaxies reflects on the average populations of formed star clusters.
In Figure~\ref{fig:density}, we show the R$_{\rm eff}$ versus the stellar surface density of our sample (colored symbols) as well as YSCs in local galaxies \citep[coral contours][]{brown2021}, Milky Way's GCs \citep[light-blue contours][]{baumgardt2018}, NSCs (black dots \citealp{Neumayer2020}), and the average densities of $z\sim8$ galaxies with masses between $10^7 - 10^8$\msun\ extracted from the relation by \citet{Morishita2024} as grey a band. We also include, critical stellar densities values (dark and light blue dot--dashed lines, respectively) above which intermediate mass black hole (IMBH) above 1000 \msun\, are predicted to form via runaway stellar collisions using the models from \citep{Rantala2026}. The two extrapolated values correspond to $2\times10^5$ \msun/pc$^2$ for YSCs forming via monolitic collapse and a factor of 10 less dens for stellar systems forming via hierarchical accretion at metallicities below 10\% Z$_{\odot}$ \citep[see][]{Rantala2025, Rantala2026}. A fair number of systems have stellar densities above $10^4$ \msun/pc$^2$ where stellar runaway collisions will favour the formation of IMBH (further discussed in Section~\ref{sec:discussion3}). Several of the smallest systems (R$_{\rm eff}<10$pc) have only upper limits on radii and thus lower limits on stellar densities. If resolved these systems will move to the top left side of the plot along the dashed lines for a given stellar mass. In general, magnifications above 50 are needed to resolve structures down to few parsec. The Cosmic Gems clusters, remain among the densest and most compact high-z star clusters. They are also among the systems with the highest magnifications (top right panel of Figure~\ref{fig:density}). We expect star clusters to loose significant mass increasing their sizes as they age, thus moving toward the region where GCs currently sit. We see that a large fraction of the currently detected high-z star cluster candidates have sizes and mass that better overlap with the NSC population detected at $z=0$. This overlap is quite interesting as mergers of star clusters is a viable channel for proto-NSC formation \citep[e.g, see review by][]{Neumayer2020}. 
The bottom left plot shows a mixed population with a broad age distribution. Detections below the $10^6$ \msun\ line becomes progressively less complete and only very young systems (age$<20$ Myr) are detected. The bottom right plot shows the dynamical age or degree of boundness ($\Pi$) of the clusters \citep[estimated by comparing their crossing time to the their ages, see][]{gieles2011, claeyssens2023}. The large majority of the star clusters are consistent with being gravitationally bound ($\rm log({\Pi})>1)$), i.e., their crossing time is significantly shorter (between 10 and 100 times) than their current stellar age \citep{claeyssens2023}. Their likelihood of surviving as coherent structures for a significant fraction of their galaxy assembly history is thus high \citep[e.g.][]{adamo2023}.  

\subsection{Star cluster mass function}
\label{sec:sc_mass_function}

\begin{figure*}
\centering
    \includegraphics[width=6cm]{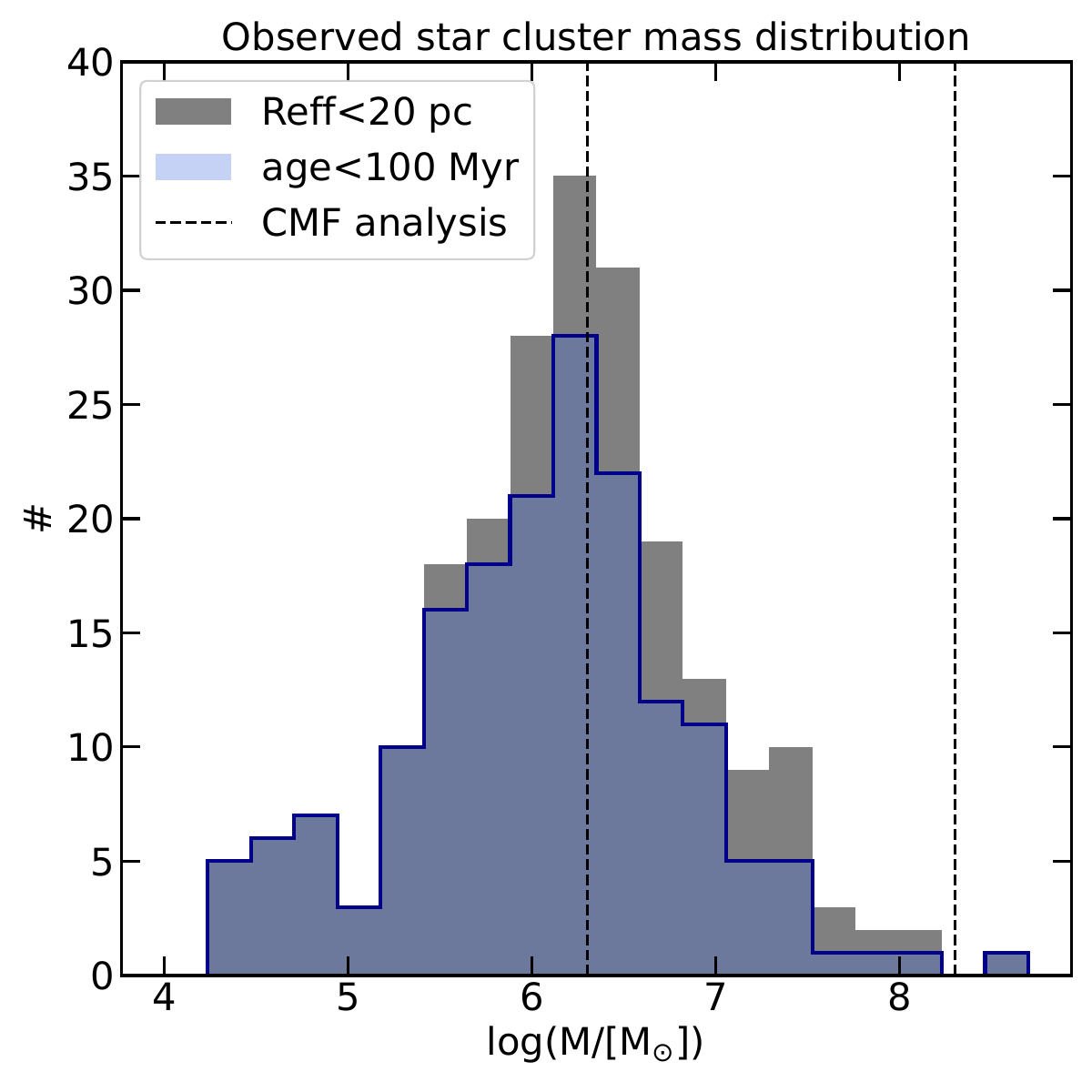}
    \includegraphics[width=6cm]{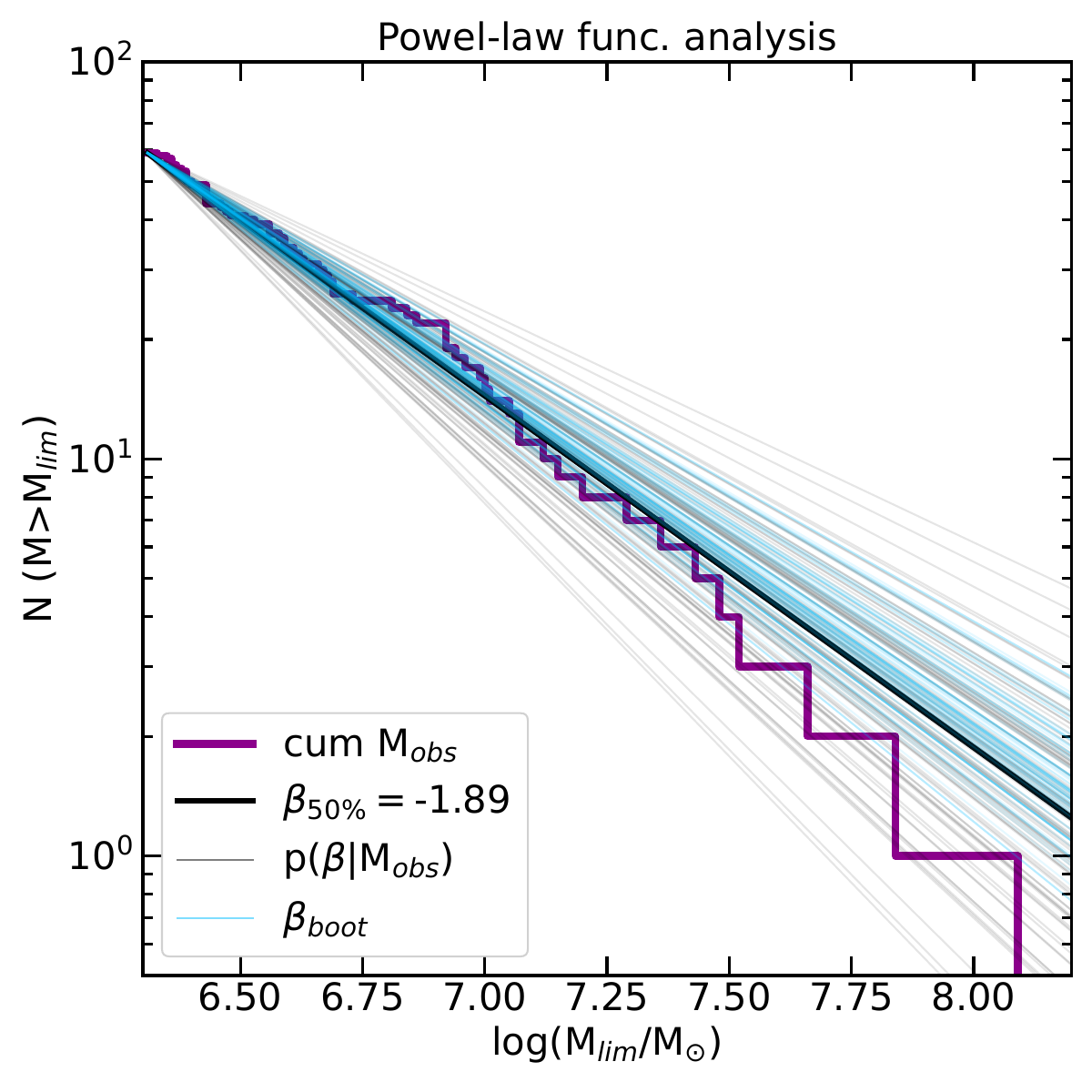}
    \includegraphics[width=6cm]{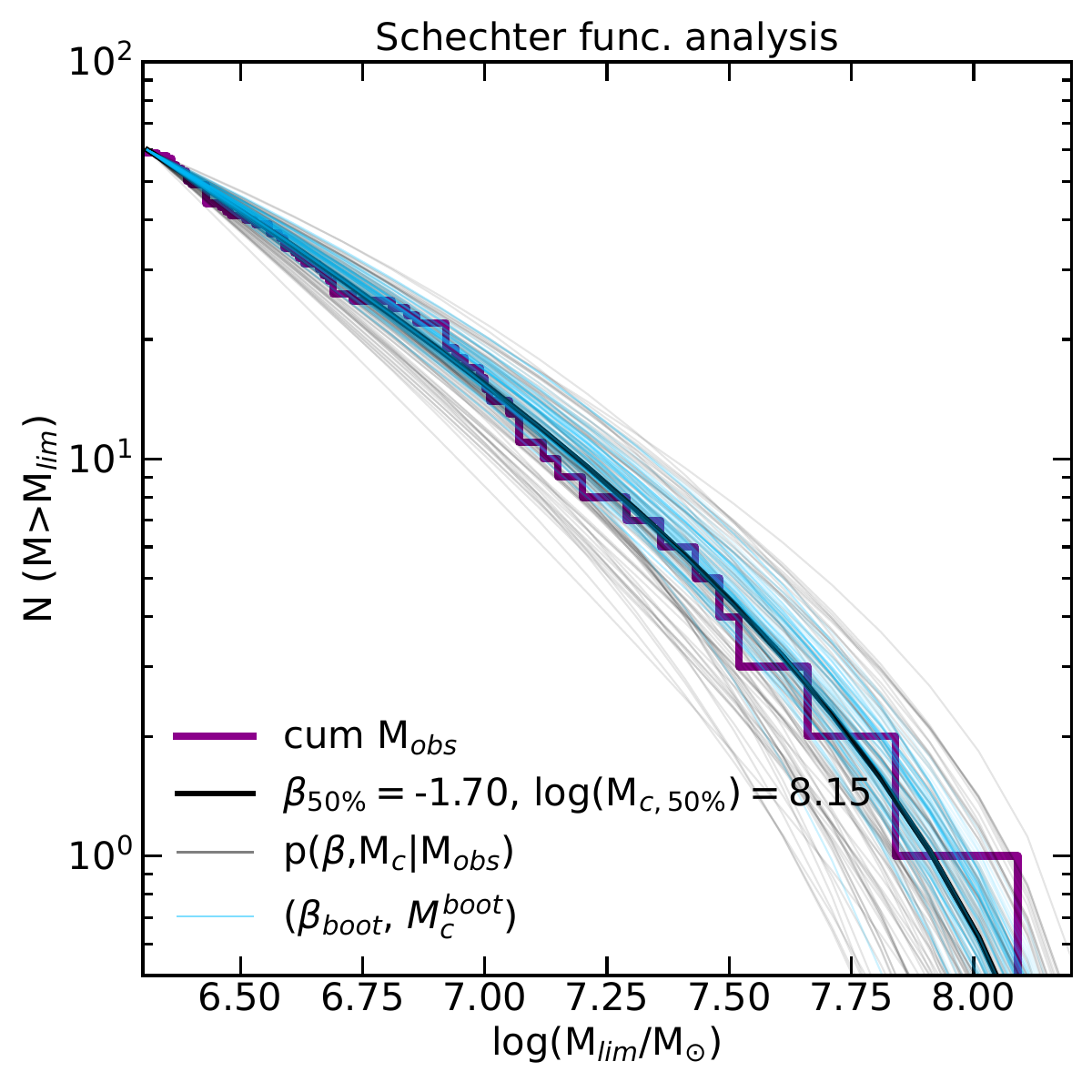}\\
	\includegraphics[width=6.5cm]{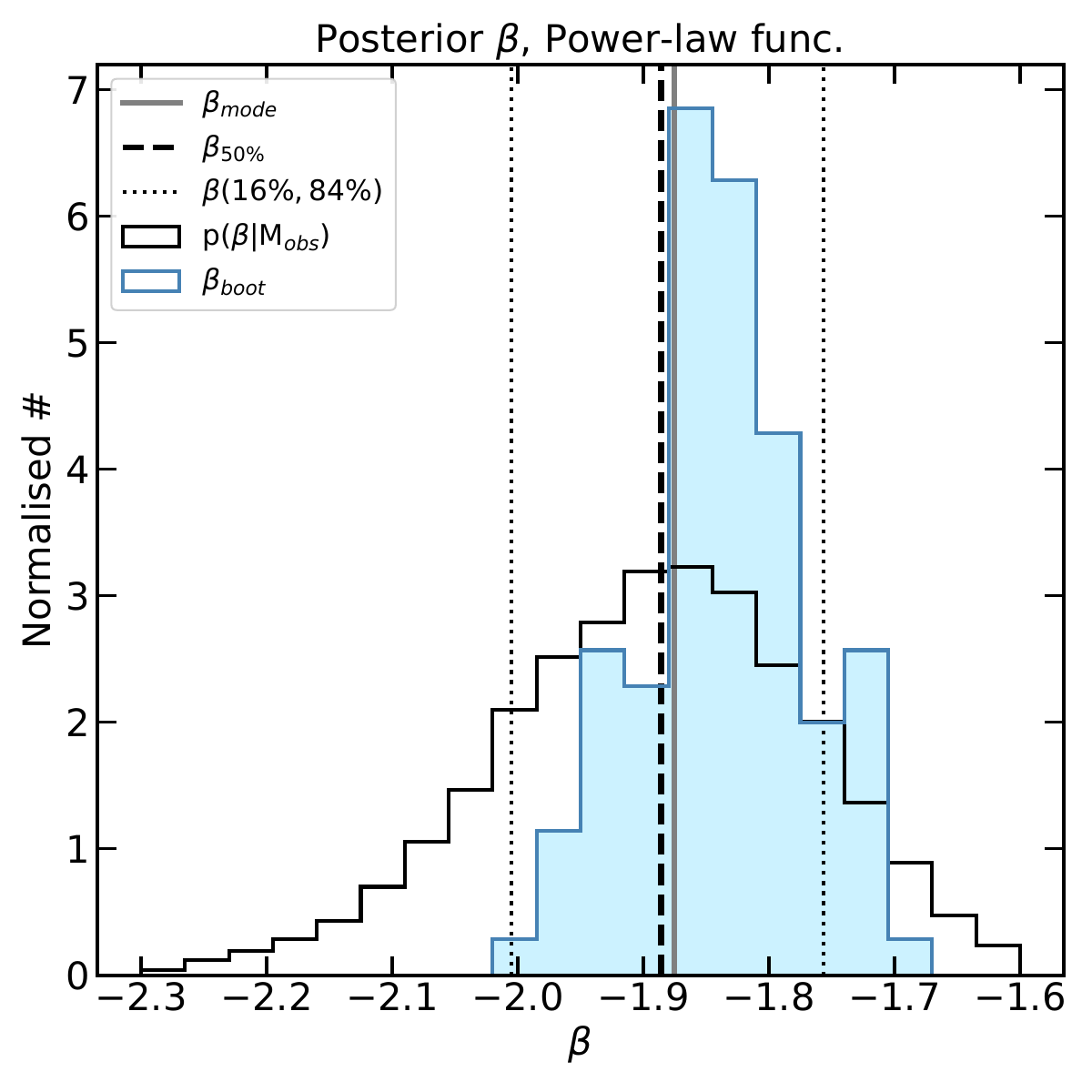}\includegraphics[width=11.4cm]{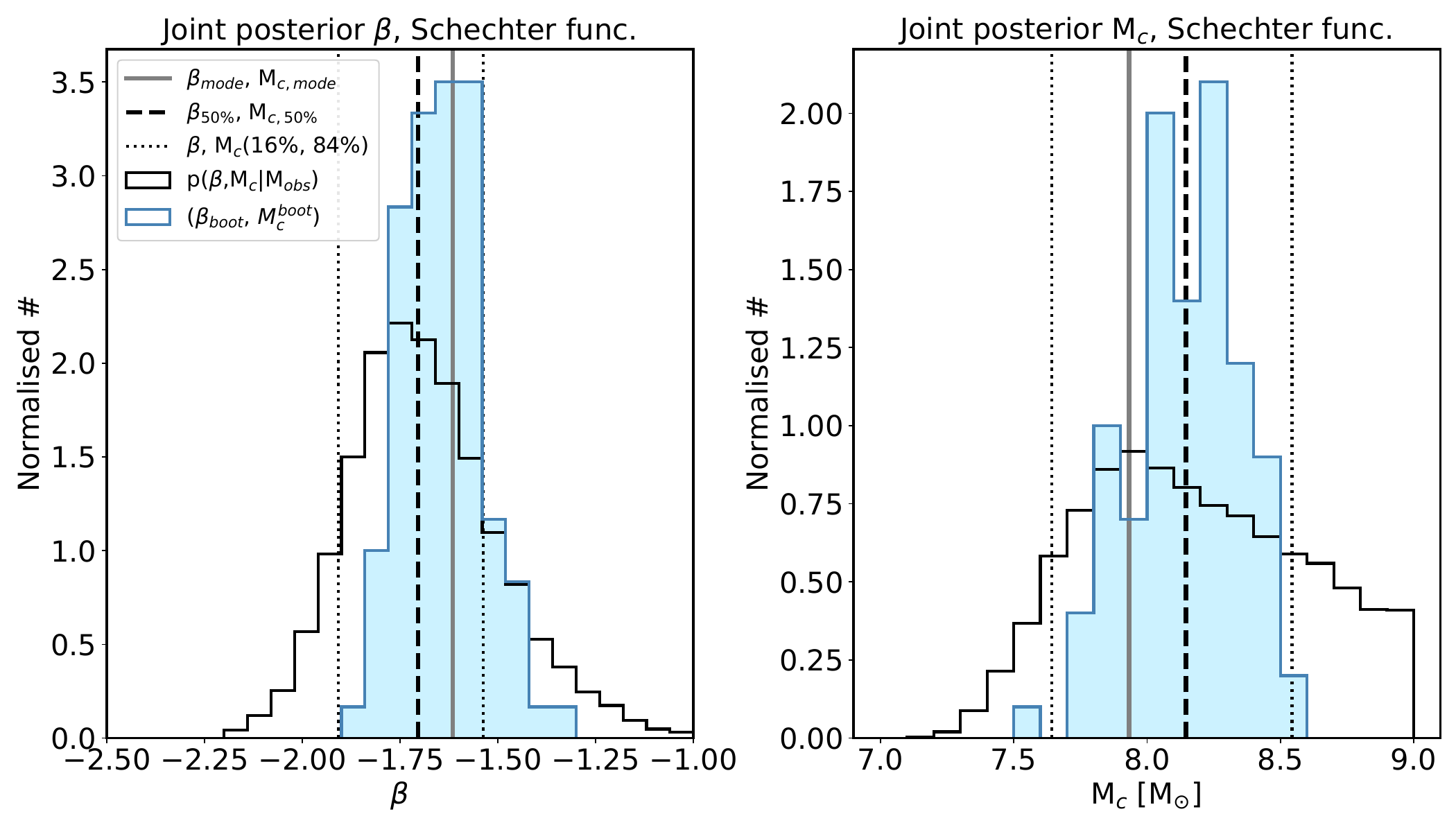}
    \caption{Star cluster mass function analysis. Top left: the grey histogram shows the mass distributions of the 222 star cluster candidates, while the dark blue one shows those with ages below 100 Myr. The vertical dashed lines indicate the low and high mass interval over which our analysis is performed and plotted as the purple line cumulative mass distributions in the top centre and right plots. Top centre and right: Observed cumulative mass distribution (purple) and reproduced distributions from the Bayesian inference analysis assuming a power-law function (top centre) nd a Schecheter function (top right). The black solid line in these two panels reproduces the mass distribution using the median values of $\beta$ (centre) and combined $\beta$ and M$_{c}$ (right) of the respective posterior distributions (bottom panels). Hundred solutions sampled from the posteriors are visualised as thin grey lines, while the light blue lines samples recovered values after bootstrapping age and mass errors.  They encompass the observed and reproduced CMF for both functions. Bottom left: Recovered posterior distribution of $\beta$ assuming a power-law function (black histogram). Mode, median, and 68\% confidence intervals are indicated as vertical solid, dashed, dotted lines, respectively. The light blue histograms shows the median $\beta$ values recovered with 500 bootstraps of the age and mass errors. Bottom centre and right: same as left panel but for the combined $\beta$ and M$_{c}$ posterior distributions (black histograms) recovered assuming a Schechter function instead.}
    \label{fig:CMF}
\end{figure*}

To build the star cluster mass function (CMF) we select among the cluster candidates, those that are younger than 100 Myr and more massive than M$_{min} = 2\times10^6$M$_{\odot}$. The top left panel of Figure~\ref{fig:CMF} shows that the age limitation does not fundamentally change the mass range of the sample. The age limit is imposed to ensure that the shape of the mass function is closest to the \emph{initial} cluster mass function and not affected by cluster disruption \citep[e.g.][]{PZ_2010review}. The minimum mass limit is selected to be close to the peak of the cluster mass distribution to ensures that incompleteness has minimum impact on the recovered slopes. Finally, we excluded from the fit the most massive cluster candidate ($\sim3$ times more massive than the second most massive) because of the low number statistics we want to avoid outliers to affect the analysis. The inclusion of this very massive system does not produce noticeable changes in the fit. We use as reference sample the one assuming a SFH with $\tau = 1$ Myr and lensing model M1. In total 60 (59 and 60, for the M2 and M3 lens models) star cluster candidates satisfy our selection, with 65\% detected in the GLIMPSE field.

We do not fit binned distributions to avoid the shortcomings of inhomogeneous weighting at different mass ranges combined with small number statistics, but use Bayesian inference to find the most probable set of values that reproduce the observed mass distribution. In the local Universe, the cluster mass function \citep[e.g.][]{adamo2020} is typically fitted by a power-law distribution as well as a power-law distribution with an exponential truncation at the high-mass end, i.e. Schechter mass function \citep{schechter1996}.  We follow here the same method introduced by \citet{johnson2017, messa2018} where the function $p_{cl}(M | \vec{\theta})\equiv\frac{p_{MF}(M | \vec{\theta})}{Z}$ describes the likelihood of finding a cluster with mass $M$ for a cluster mass function $p_{MF}(M | \vec{\theta})$ and normalization $Z$. The mass distribution is described by the two following functions:
\begin{equation}
\label{eq:bay_trun}
p_{MF,sch}(M | \vec{\theta})\propto M^{\beta}e^{-M/Mc}\ \Theta(M_{min}) 
\end{equation}
for the Schechter one (where the slope $\beta$ and the characteristic mass M$_c$ are the parameters), and 
\begin{equation}
\label{eq:bay_untrun}
p_{MF,pl}(M | \vec{\theta})\propto M^{\beta}\ \Theta(M_{min}) 
\end{equation}
for the power-law one (where $\beta$ is the only parameter).
In both cases we limited the study of the mass function to masses above $M_{min} = 2\times10^6$ \msun. 
We use Bayes' theorem to derive the posterior probability distribution function of the parameters $\vec{\theta}$, defined as:
\begin{equation}
p(\vec{\theta} | \{M_{obs}\})\propto p_{cl}(\{M_{obs}\} | \vec{\theta})p(\vec{\theta}),
\end{equation}
where $\{M_{obs}\}$ is the observed mass distribution and $p(\vec{\theta})$ is the prior probability of the parameters $\vec{\theta}$. We choose a flat prior probability distribution to cover the range of possible values $-3<\beta<-1$ and $\rm{log}(M_{min}/M_\odot)<\rm{log}(Mc/M_\odot)<9$.

The posterior probability distributions have been sampled with the \texttt{Python} package \texttt{emcee} \citep{emcee}, by implementing a Markov Chain Monte Carlo (MCMC) sampler from \citet{emcee2010}. We used 100 walkers, each producing 600 step chains. After discarding the first 100 burn-in steps of each walker, we recovered 50000 independent sampling values for each fit. We report in Table~\ref{tab:CMF}, the median and 16 and 84\% uncertanties of the posterior distributions for each fitted parameter: $\beta$ for a power-law and $\beta$,M$_c$ for a Schechter function. Using the \texttt{scipy.stats.gaussian\_kde} package, we estimated the mode of the posterior distributions, the latter also reported in the Table.  Discrepancies between mode and median values are usually interpreted as a lack of convergence of the fit.  

We test the robustness of the derived parameters by analysing the mass distributions obtained by using two other independent lens models. The recovered parameters for power-law and Schechter functions are included in Table~\ref{tab:CMF}.

Finally, we test the recovered CMF parameters by bootstrapping the uncertainties on masses and ages of the star clusters. We randomly sampled 500 populations and applied the selection criteria above ($6.3<log(M)<8.2$, age$<100$ Myr). For each realisation of a population we derived median values of the parameters from the posterior distributions. 

In Figure~\ref{fig:CMF} we summarise the outcome of the CMF analysis. For visualisation purposes we plot the observed cluster mass function as a cumulative distribution (purple step line in top centre and right panels). In the top central and right panels, the black thick line shows the power law mass distribution with slope $\beta = \beta_{50\%} = -1.89$ (centre) and Schechter function with $\beta_{50\%} = -1.70$ and characteristic mass, log(M$_{c,50\%})=8.14$ (right) sampled with the same number of objects as in the observed cumulative distributions. The thin-grey solid lines show 100 realisations extracted from the posterior distributions (black histograms, bottom panels). The light blue thin lines show 100 random solutions recovered from the bootstrapping when mass and age errors are propagated (light blue histograms, bottom panels). The analysis of the observed CMF derived with other lensing models (see Table~\ref{tab:CMF}), produces very similar slopes, suggesting that overall the recovered mass distributions are consistent with each others.

The bottom left panel of Figure~\ref{fig:CMF} shows the recovered posterior distribution of the observed CMF, including median, mode, and 16, 84 \% intervals. Overall, we see a convergence of the fit to a slope $\beta_{50\%} = -1.89^{+0.13}_{-0.12}$. Median and mode are well within the 1$\sigma$ confidence level.  The bootstrapping analysis produces $\beta$ distributions (light blue histograms in bottom panels)  slightly offset from the median values but overall within 1 $\sigma$. An inspection of the error estimates shows that this behaviour is likely due to the fact that the errors are not symmetric around the best values, but skewed toward larger values. On the bottom centre and right side of Figure~\ref{fig:CMF}, we show similar plots, but for the joint--posterior distributions of slope and characteristic mass, $\beta$ and M$_c$, when a Schechter function analysis is performed (see Appendix A for a 2D visualisation of the posteriors). The latter functional form reproduces closely the observed mass distribution (top right panel). We find $\beta_{50\%} = -1.70^{+0.17}_{-0.20}$ and a characteristic mass, log(M$_{c,50\%})=8.14^{+0.40}_{-0.50}$. In the bottom right plot, it is possible to notice that the posterior distribution of M$_{c}$ (black histogram) is not exactly symmetric and that only 2 clusters have masses closed to M$_{c,50\%}$. The lack of full convergence makes the presence of a truncation on the cluster mass function not statistically significant. The bootstrapping analysis confirms the recovered values with uncertainties. The recovered slope is within 1 $\sigma$ in agreement with the power-law fit value. We conclude that the CMF obtained by combining YSC detected between redshift 1 and 10 is consistent with a power-law mass function of slope close to $-2$. We do not have enough statistics to quantify the presence of a exponential truncation. We will discuss these results in the next section.

\begin{table*} 
\centering
\begin{tabular}{l|cc|cccc} 
\hline
Lens Mod & \multicolumn{2}{@{}c@{}}{Power Law Function} &  \multicolumn{4}{@{}c@{}}{Schechter Function} \\

& $\beta_{mode}$ & $\beta_{50\%}$ &  $\beta_{mode}$ & $\beta_{50\%}$ & log(M$_{c, mode}$/[M$_{\odot}$]) & log(M$_{c, 50\%}$/[M$_{\odot}$])\\
\hline
M1 (ref) & $-1.87$ & $-1.89_{-0.12}^{+0.13}$ &  $-1.61$ &$-1.70_{-0.20}^{+ 0.17}$& $7.93$ & $ 8.14_{-0.50}^{+0.40}$\\

M2 & $-1.87$ & $-1.89_{-0.12}^{+0.13}$ & $-1.72$ &$-1.71_{-0.20}^{+ 0.16}$& $8.20$ & $ 8.15_{-0.48}^{+0.40}$\\

M3 & $-1.86$ & $-1.88_{-0.11}^{+0.13}$ &  $-1.62$ &$-1.65_{-0.23}^{+ 0.19}$& $7.87$ & $ 7.97_{-0.52}^{+0.34}$\\
\hline
\end{tabular} 
\caption{Modes and medians of the posterior distributions recovered from the analysis of the observed CMF assuming a power-law distribution and a exponentially truncated power-law (Schechter) function. See main text for derivation and discussion of these values. The values are given for the three different lens models.}
\label{tab:CMF} 
\end{table*}

\section{Discussion}
\label{sec:section5}

\subsection{Observing Star clusters at high redshift}
\label{sec:discussion1}

The identification of star cluster candidates at high redshift remains a significant observational challenge, primarily due to the limited spatial resolution (i.e., it needs high lensing magnification to reach parsec scales with the current telescopes) and the very few physical information accessible (most of them being only detectable in deep photometric data). The observed compact stellar clumps, extracted from NIRCam observations of magnified galaxies at $1<z<10$, typically exhibit effective radii on the order of tens of parsecs after PSF deconvolution and magnification correction, with only a small fraction of them (between $4$ to $20$ \%, depending on the samples) being consistent with the expected scales of young and massive star clusters in local Universe. Nonetheless, blending, intrinsic limits of the spatial resolution and surface brightness dimming introduce ambiguities in the selection of star cluster candidates, since only based on empirical compactness and size criteria. 

\noindent In order to isolate star cluster candidates out of the total clumps sample, we applied a unique size threshold, fixed at $20$ pc. This threshold is arbitrary and has been chosen to be similar to the limit used for local Universe star cluster studies (\citealt{Linden2017,adamo2020}). In our total sample, among the 2540 extracted stellar clumps, 19\% of them remain unresolved with respect to the local PSF. 
We include these clumps in the star cluster sample if the size upper limit value is below 20 pc. In total, we gathered 222 star cluster candidates. If we set the selection threshold at 30 or 50 pc, we would have selected 353 and 723 cluster candidates, respectively, with 40 to 50\% of then being unresolved, thus being probably smaller, but with the risk of including also a large quantity of low-density ($<10^{2} \ \rm M_{\odot}/pc2$) and high mass ($>10^8 \ \rm M_{\odot}$) structures sharing similar properties to high-redshift JWST individual galaxies (\citealt{Morishita2024}, 13\% of their $z>5$ galaxies exhibit sizes $<100$ pc). By selecting clumps with size below 20 pc, we select only very compact and dense structures (with the majority of them denser than $10^{2} \ \rm M_{\odot}/pc^2$). While there is a non-zero probability that not all the selected star cluster candidates are exactly individual systems, it is likely that the physical properties we derive are dominated by the brightest (most massive) star cluster within the clump.

\subsection{Star cluster formation redshift}
\label{sec:discussion2}

The observation of strong lensing clusters with JWST/NIRCam opened a new window on the study of star clusters in the distant Universe. On one hand, the detection of very high redshift ($z>6$) star clusters with JWST (\citealt{vanzella2023, adamo2024, mowla2024}) potentially suggest that star clusters might be forming in the bursty events \citep{vanzella2025} abundantly recovered in early galaxies \citep{matthee2025}, with very high star formation efficiency as reported by simulations (see \citealt{Pascale2025, Polak2024, Menon2025} for link between star cluster formation and star-formation burst). Hence, star clusters might be strong candidates to significantly contribute to the reionisation of the Universe (see \citealt{Ricotti2002, SC2011, Katz2013, KatzR2014, BK2018, Ma2021} for discussion on the contribution of star cluster to the reionisation). On the other hand, the detection of CN GCs ($>1$ Gyr old), such as the Sparkler galaxy clusters (\citealt{Mowla2022, claeyssens2023, adamo2023,Whitaker2025}), suggests that GCs can form at $z>6$ and survive through cosmic time at least down to $z\sim 1$ where they are currently detected and potentially down to $z\sim 0$. 

\noindent Figure~\ref{fig:formation_redshift} shows that the majority of  star clusters in our sample (72\%) have formed at the CN ($1<z<4$), we obtain very similar proportion if we include also clumps with size between 20 and 30 pc.  These clusters span a wide range of surface stellar densities, from $10^1$ to $10^5 \ \rm M_{\odot}/pc^2$. Among the sample, we detect 35 objects observed at $z>6$ (with 11 of them being from the Cosmic Gems and Sunrise arcs), as well as 11 star clusters observed at CN but formed at $z>4$. All these clusters share similar densities (systematically higher than $10^3 \ \rm M_{\odot}/pc^2$). The lack of low-density and/or old star clusters is likely due to an observational bias (because of surface brightness dimming effects with redshift as well as fainter luminosities for older age clusters). If they survive down to $z=0$, these star clusters formed around $z=8 - 12$, would reach the age of $\sim 12 - 13$ Gyrs.
Multiple numerical simulations of star cluster formation and evolution across redshift have proposed that denser clusters meet the physical conditions required for long-term survival into the GC regime and thus, have more chances to survive down to $z=0$ and become the progenitors of GCs (\citealt{Pfeffer2018,Pascale2025,Calura2025,williams2025,Rodriguez2022,Rodriguez2023}). Simulations agree that a large fraction of Milky Way GCs formed at CN. It remains unclear what the fate of the oldest star clusters formed at redshift above 8 and beyond is. Improved recipes for cluster formation and better understanding of the observational properties of the first star clusters will provide tighter constraints on the formation and evolution of the first star clusters in the future.

\subsection{High-redshift star cluster densities}
\label{sec:discussion3}

In Section~\ref{sec:density_size}, we report that the stellar densities and sizes of the gathered cluster candidates  typically exceed the densities of GCs and better overlaps with NSCs. Definitively, incompleteness affects the recovery of less massive (fainter) systems. It is also likely that the most extended objects between 10 and 20 pc could be blends of multiple star clusters. However, we remark that a change of a factor of few in mass is less relevant than a change of a factor of 10 is size, because of the quadratic inverse dependence from the radius, resulting in similar or higher densities. 

\noindent The stellar density of star clusters and clumps has attracted large attention in the theoretical community. Caution, however, is necessary when comparing observations and simulations. Even for the state-of-the-art simulations it remains a challenge to measure stellar densities of stellar aggregates (star clusters and stellar clumps) due to the necessity to implement a softening radius (even of parsec scales) which strongly affects the resulting densities \citep{lahen2025b}. It is therefore necessary to resolve not only single stars but their N-body interactions to achieve robust measurements of stellar densities \citep{lahen2025a}. Simulations of idealised isolated molecular clouds find that very compact, bound, massive star clusters form with high star formation efficiency ($\rm SFE>80$\%) and extremely short time scales (about 1 free-fall time, that is $<2$Myr), reaching stellar densities of $10^3$-$10^4$ \msun/pc$^2$ \citep{fukushima2023, Polak2024, fujii2024}. When pushing to the limits of what computational CPU allows, it is now possible to resolve massive cluster formation and the host galaxy star-by-star \citep{lahen2025b}. Thanks to the presence of a surrounding host environment, massive star clusters form out of a fast hierarchical growth (mergers of sub-clusters forming with very high star formation efficiencies, compact and very short free-fall times) and reach stellar densities of $10^5 \rm \ M_{\odot}/pc^2$  \citep{lahen2025b} comparable to the stellar surface densities we observe in high-redshift galaxies. A different approach is to evolve galaxies in a cosmological setting pushing to softening radii of a few parsecs but following self-consistently cluster formation and evolution along the host galaxy assembly history \citep{Calura2025, mayer2025, williams2025}. While these simulations cannot reproduce the stellar densities measured at the smallest physical scales in observations, they can reproduce the distributions of stellar densities we obtain for systems with sizes of 10-20 pc, suggesting that we are not too far from a convergence.

Finally, it is important to highlight the implications that such observed stellar densities have for IMBH seed formation and growth. JWST observations have opened an unprecedented window into the earliest galaxies, showing that their compactness and abundances point toward star clusters being a major mode of star formation within them \citep{naidu2025}. N-body simulations of star clusters show that at stellar densities above $10^4$ \msun/pc$^2$ stellar runaway collisions in the core of very dense stellar systems are a viable channel to grow IMBH \citep[e.g][among many others]{katz2015, arcasedda2024, fujii2024, Rantala2025, vergara2025} but the rate of retention of such BHs remain uncertain \citep{Rantala2024}. The metallicity is an important factor for the resulting IMBH formation efficiency which decrease at higher metallicity due to the change in wind mass loss rates as reported by \citet{Rantala2026}. Another important factor is the mechanism driving star cluster formation. If clusters form via hierarchical accretion of denser sub-clusters, then the critical stellar density of the resulting star cluster forming an IMBH can be lower because sub-cluster mergers will effectively increase the final radius of the resulting cluster \citep{Rantala2025}. Our measurements show that a good number of star clusters candidates already reach stellar densities above $\sim10^4$ \msun/pc$^2$, but their fraction is likely larger (because of upper-limits and possibly effects of blending affecting the current selection, i.e. $<$20 pc). Next generation ELT instruments such as MICADO, will play a fundamental role in enabling a better characterisation of the intrinsic sizes of these star cluster candidates necessary to derive tighter constraints on their stellar densities. It is therefore important to continue explore stellar densities of large samples of star clusters and clumps to gain further insights into the galaxy conditions that lead to the formation of highly dense stellar systems.

\subsection{The cluster mass function}
\label{sec:mass_function}

We have combined the cluster candidates selected between redshift 1 and 10 to derive the first direct measurement of the cluster mass function. This is a key cosmic time for the formation of star clusters that will then evolve and in some fractions survive to become nowadays GCs. In the local Universe, YSCs typically populate power-law mass distributions with slope $\sim-2$ \citep{krumholz2019}. It remains still under debate whether the upper-mass end of the CMF is better fitted by an exponential truncation, M$_c$, in the form of a Schechter function \citep{adamo2020}. The presence of a truncation and in particular its variation as a function of SFR density of the galaxy \citep{Wainer2022}, would suggest a tight correlation between the forming population of star clusters, their maximum mass and the global physical properties of the host galaxies, represented by the SFR density \citep{adamo2020}. On the other hand, it is well known that the GC mass function measured in the local Universe has an approximate log-normal shape with a peak (turn-over) at $\sim2\times10^5$M$_{\odot}$, which appears to be a constant of the GC populations \citep{brodie2006}. The origin of the nearly-constant mass (luminosity) turn-over of GC populations remains still unknown, since internal mechanisms such as evaporation and mass loss due to stellar evolution are not sufficient to account for its almost universality \citep[e.g.][]{fall2001,kruijssen2009}. Even for the GC mass function it has been reported the presence of a truncation at the high-mass end of the distribution varying as a function of host galaxy mass in the form of a Schechter characteristic mass \citep{jordan2007}. It remains unclear whether that truncation is an imprint left at birth or due to the long evolution GCs have been through. 
With this study we are taking the first steps into bridging the cluster formation and evolution across cosmic time. Our first direct measurement of the CMF suggests that even at high redshift, when galaxies are still assembling the majority of their mass, YSCs have a mass function that is compatible with a power law distribution with slope $-2$ in the mass range $2\times10^6$--$1\times10^8$ M$_{\odot}$. Even assuming that these YSCs will lose 50\% of their initial mass by redshift 0, we are still beyond the reach of probing star clusters that would constitute the peak of the GC mass function. It has been recently reported that in low mass galaxies like the Cosmic Gems \citep{adamo2024} and the Firefly Sparkler \citep{mowla2024}, there might not be enough mass in the host galaxies to actually populate a cluster mass function down to thousand solar masses \citep{vanzella2025}. Under the reasonable assumption of high cluster formation efficiency, \citet{vanzella2025} conclude that either a shallower power law or a high-mass limit (i.e., not low mass clusters can form) need to be advocated to reconcile cluster formation in these early galaxies. Our study provides one more piece to the puzzle, suggesting that the power-law $-2$ function appears to be in place even at high redshift, thus, pointing to a higher efficiency of massive cluster formation and a change in the physics of cluster formation in general. We conclude that early galaxies do not sample clusters bottom-up with low masses first as expected by the stochastic sampling of a power-law $-2$ distribution. In this context, the investigation of a Schechter characteristic mass becomes  relevant as it could describe the maximum cluster mass a galaxy can form. However, with 60 clusters over a large redshift bin, there is not much room for investigation. We do not find a statistically significant signal of the presence of a truncation in the mass function but we cannot rule out that it is there. Understanding how clusters form and evolve along with their host galaxies remains still a pressing question. We need deep JWST surveys like GLIMPSE to gather large and complete samples of star cluster populations reaching $10^5$ M$_{\odot}$ and hopefully lower to build a model that can describe them self-consistently.  



\section{Conclusions}

In this work, we have presented the first large sample of high-redshift ($z>0.5$) star clusters, identified in JWST/NIRCam imaging of six lensing fields. By combining the ultra-deep images of the lensing cluster AbellS1063 from the GLIMPSE program with complementary literature samples, we gathered 222 compact stellar systems exhibiting effective radius smaller than 20 pc, reaching absolute magnitude (in the V band, rest-frame) of $-10$. 
We performed cluster PSF-like photometry in the available JWST/NIRCam bands as well as SED fitting with BAGPIPES.  
Our results highlight the transformative role of JWST in opening a new window onto star cluster formation across cosmic time, and the need of very deep observations (reaching AB magnitude $<$ 30 in point-source objects), combined with high lensing magnification ($\mu>10$) to detect such systems that would otherwise remain inaccessible. 
With the current sample, we show that: 
\begin{itemize}
    \item The majority of the star clusters are young (with ages $<100$ Myr, cf Figures~\ref{fig:histo_sample} and \ref{fig:formation_redshift}). Their formation redshift peaks at CN ($1<z<4$), but about 15\% form in reionization era. We find a small fraction of evolved systems (with age $>1$ Gyr) at CN with formation redshift prior to $z~4$, providing a direct links between high-z star clusters and GC populations already in place at CN.
    \item The physical properties of the high-redshift star clusters (high stellar densities with $\rm \Sigma_{M_*}>10^2 \ M_{\odot}/pc^2$, compact sizes and stellar masses ranging from $\rm 10^4$ to $10^8 \ \rm M_{\odot}$) exceed those of local star clusters suggesting that they could be counterparts of the local GCs. Their elevated stellar densities make them an ideal channel for runaway stellar collisions and intermediate mass black hole growth. 
    \item We provide the first direct measurement of the young star cluster mass function at high-redshift, based on 60 objects (cf Sections~\ref{sec:mass_function},\ref{sec:discussion3}), recovering a power-law slope close to the canonical -2, measured in the local universe. It remains unexplored whether there is a possible evolution of the cluster mass function with redshift and galaxy physical properties. 
\end{itemize}
Future work combining a larger sample of deep JWST imaging, spectroscopic follow-up, and comparisons with high-resolution simulations will be essential to constrain the survivability of these clusters and to establish their role in the assembly of galaxies and the origin of present-day GC populations.

\begin{acknowledgements}
AA thanks Antti Rantala for interesting discussions about IMBH formation. AA acknowledges support from the Kunt \& Alice Wallenberg Foundation under grant ID 2024.0110, the Swedish research council Vetenskapsr{\aa}det (VR) project 2021-05559 and VR consolidator grant 2024-02061. The Dunlap Institute is funded through an endowment established by the David Dunlap family and the University of Toronto. We acknowledge the support of the Canadian Space Agency (CSA) [25JWGO4A06]. HA acknowledges support from CNES, focused on the JWST mission and support by the French National Research Agency (ANR) under grant ANR-21-CE31-0838. 

This work is based on observations obtained with the NASA/ESA/CSA \textit{JWST} and the NASA/ESA \textit{Hubble Space Telescope} (HST), retrieved from the \texttt{Mikulski Archive for Space Telescopes} (\texttt{MAST}) at the \textit{Space Telescope Science Institute} (STScI). STScI is operated by the Association of Universities for Research in Astronomy, Inc. under NASA contract NAS 5-26555. 
Support for program \#3293 was provided by NASA through a grant from the Space Telescope Science Institute, which is operated by the Association of Universities for Research in Astronomy, Inc., under NASA contract NAS 5-03127.      
\end{acknowledgements}

%
%


\bibliographystyle{aa} 
\bibliography{aanda}  

\begin{appendix} 

\section{Joint-posterior distributions recovered from the Schechter mass function analysis}

\begin{figure}
	\includegraphics[width=9cm]{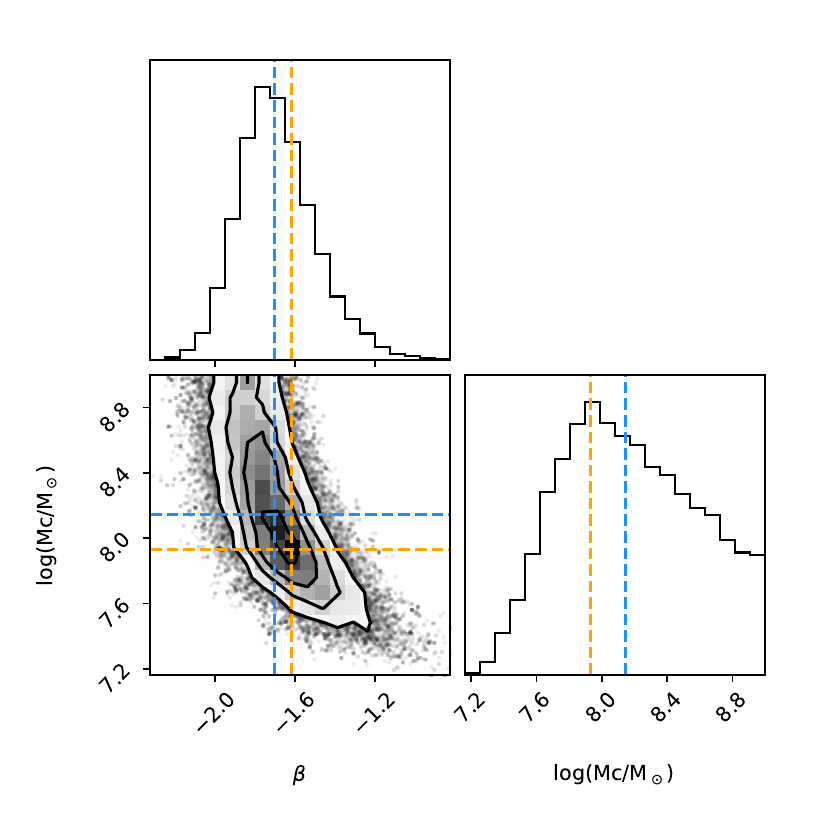}
    \caption{Joint-posterior distributions for the $\beta$ slope and M$_c$ parameters, using the cluster catalogue produced by the reference lens model. The most-likely (mode) and median values are visualised by orange and blue dashed lines respectively.}
    \label{fig:posterios_schfunc}
\end{figure}

In Figure~\ref{fig:posterios_schfunc}, we display the joint-posterior distributions for the analysis of $\beta$ and M$_c$ presented in Section~\ref{sec:mass_function}. The plotted $\beta$ and M$_c$ distributions are also showed in Figure~\ref{fig:CMF} of the main text. We plot in blue and orange dashed lines the median and mode of the recovered values in Figure~\ref{fig:posterios_schfunc}. There is degeneracy between the two values, which is enhanced by the lack of convergence in the M$_c$ parameters. See main text for detailed discussion

\section{Comparison with literature cluster properties}
We compare the physical parameters of the star cluster obtained from the\texttt{Bagpipes}SED fits (using the BPASS tables) with the values published in previous studies for the SMACS0723 sample (\citealt{claeyssens2023}), the Sunrise arc (\citealt{vanzella2023}) and the Firefly Sparkler (\citealt{mowla2024}). 

\begin{figure*}
	\includegraphics[width=18cm]{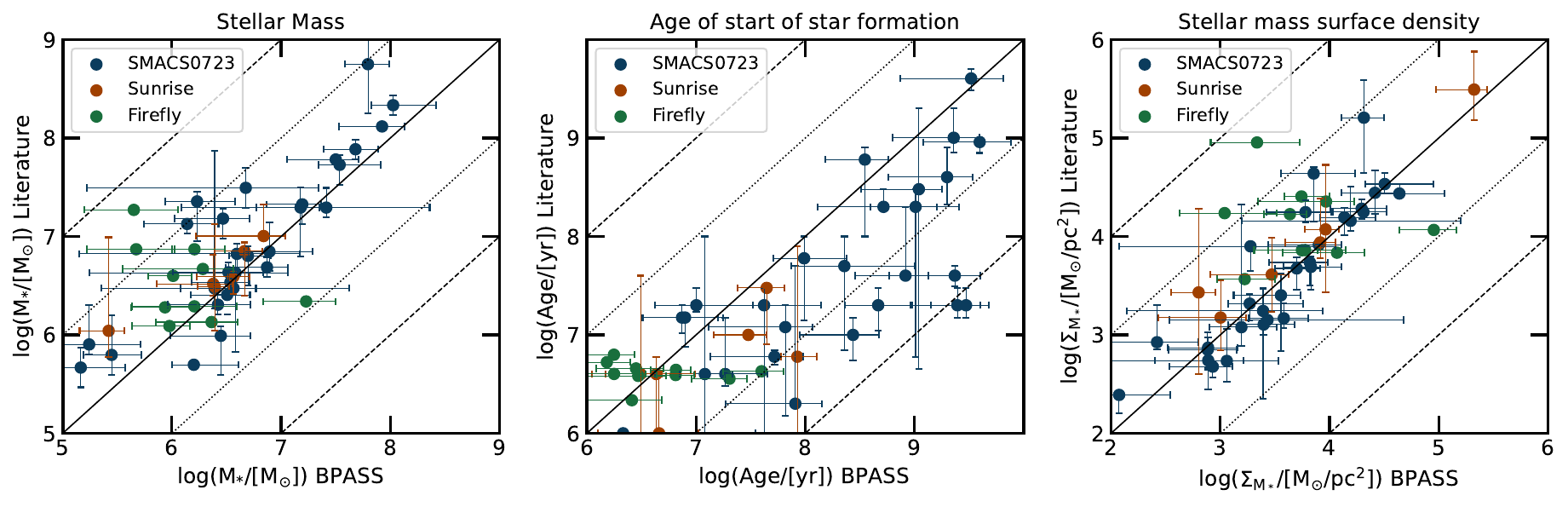}
    \caption{Comparison between the star clusters properties measured with\texttt{Bagpipes}and the properties published in the previous studies for the SMACS0723 sample (blue), the Sunrise arc (orange) and the Firefly Sparkler (green). From left to right, the panels present the stellar mass, the age of the start for star formation and the stellar mass surface density. The solid lines represent the 1:1 relation, while the dotted and dashed lines represent the $\rm 1:\pm 10$ and $\rm 1:\pm 100$ relations, respectively.}
    \label{fig:comp_lite}
\end{figure*}

\section{Comparison between three lens models for GLIMPSE}
\label{app: lensmodelcomp}
We compare the GLIMPSE star clusters properties obtained from three different lens models described in Section~\ref{sec:lens_models}.
\begin{figure*}
	\includegraphics[width=18cm]{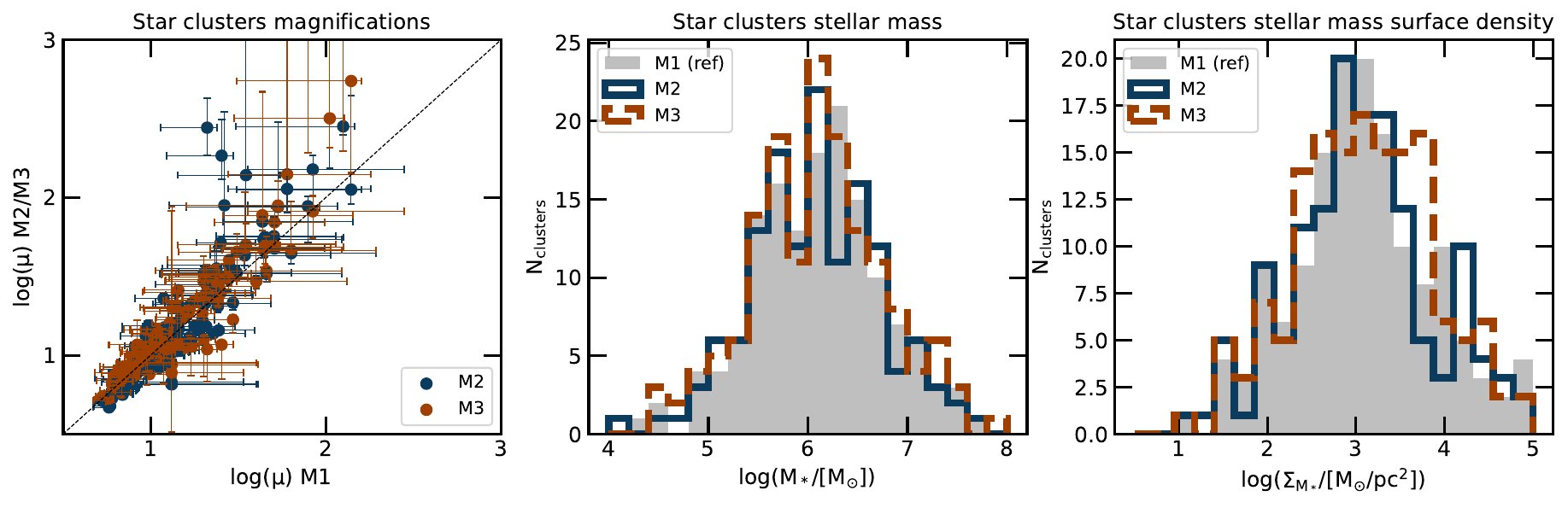}
    \caption{Comparison between the GLIMPSE star clusters properties measured with three different lens models: M1 (used as reference model), M2 and M3, as described in Section~\ref{sec:lens_models}.}
    \label{fig:comp_models}
\end{figure*}

\end{appendix}
\end{document}